\newif\ifTWO
\global\TWOtrue

\ifTWO
\documentclass[]{MyAgujournal2019}
\else
\documentclass[]{agujournal2019}
\fi

\journalname{Paleoceanography and Paleoclimatology}

\usepackage{wasysym} 
\usepackage{color}
\usepackage{graphicx}

\usepackage[]{natbib}

\graphicspath{{./}{../eps/}}

\usepackage{amsmath}

\usepackage[T1]{fontenc}
\usepackage[osf]{baskervillef} 
\usepackage[varqu,varl,var0]{inconsolata}
\usepackage[scale=.95,type1]{cabin}
\usepackage[baskerville,vvarbb]{newtxmath}
\usepackage[cal=boondoxo]{mathalfa}

\usepackage[colorlinks]{hyperref}
\hypersetup{
  citecolor=blue, 
  linkcolor=blue,
   urlcolor=blue
}

\ifTWO
\def\f1scale{0.50}
\else
\def\f1scale{0.80}
\usepackage{lineno}
\linenumbers
\fi

\newcommand{\tcr}{\textcolor{black}}

\newcommand{\noi}{\noindent}
\newcommand{\fns}{\footnotesize}
\newcommand{\scs}{\scriptsize}
\newcommand{\lsim}{\mbox{\raisebox{-.0ex}{$\stackrel{<}{_\sim} \ $}}}
\newcommand{\gsim}{\raisebox{-.5ex}{$\stackrel{>}{\sim} \ $}}
\newcommand{\bm}{\boldmath}
\newcommand{\ubm}{\unboldmath}
\newcommand{\q}{\frac}
\newcommand{\e}[1]{\mbox{$\x10^{#1}$}}
\newcommand{\x}{\times}
\newcommand{\D}{\Delta}
\newcommand{\pmo}{\mbox{$^{-1}$}}
\newcommand{\beqn}{\begin{eqnarray}}
\newcommand{\eeqn}{\end{eqnarray}}

\renewcommand{\v}[1]{\mbox{\bm$#1$\ubm}}
\newcommand{\vs}{\mbox{$\v{s}$}}
\newcommand{\vn}{\mbox{$\v{n}$}}

\newcommand{\vb}{\mbox{$\v{b}$}}

\newcommand{\vL}{\mbox{$\v{L}$}}
\newcommand{\vT}{\mbox{$\v{T}$}}
\newcommand{\bbar}{\mbox{$\overline{\v{b}}$}}
\newcommand{\pbar}{\mbox{$\bar{p}$}}
\newcommand{\ombar}{\mbox{$\bar{\omega}$}}
\newcommand{\sm}{\mbox{$\sim$}}
\renewcommand{\*}{\cdot} 

\newcommand{\alp}{\mbox{$\alpha$}}
\newcommand{\tldalp}{\mbox{$\widetilde{\alpha}$}}
\newcommand{\bet}{\mbox{$\beta$}}
\newcommand{\del}{\mbox{$\delta$}}

\newcommand{\om}{\mbox{$\omega$}}
\newcommand{\Om}{\mbox{$\Omega$}}

\newcommand{\eps}{\mbox{$\epsilon$}}
\newcommand{\kap}{\mbox{$\kappa$}}
\newcommand{\epss}{\mbox{$\epsilon^*$}}
\newcommand{\gtf}{\mbox{$g_{25}$}}
\newcommand{\gft}{\mbox{$g_{43}$}}
\newcommand{\sft}{\mbox{$s_{43}$}}

\newcommand{\gfti}{\mbox{$g_{43}^{-1}$}}
\newcommand{\sfti}{\mbox{$s_{43}^{-1}$}}
\newcommand{\sigot}{\mbox{$\sigma_{12}$}}
\newcommand{\sigft}{\mbox{$\sigma_{43}$}}
\newcommand{\tauot}{\mbox{$\tau_{12}$}}

\newcommand{\eM}{\mbox{$e_{\scs \mercury}$}}

\newcommand{\obl}{\mbox{$\epsilon$}}
\newcommand{\prc}{\mbox{$\phi$}} 
\newcommand{\vpi}{\mbox{$\varpi$}}
\newcommand{\ggp}{\mbox{$\gamma_{gp}$}}
\newcommand{\asy}{\mbox{$''$/y}}
\newcommand{\Psin}{\mbox{$\Psi_0$}}
\newcommand{\ktt}{\mbox{$k_2$}}
\newcommand{\dtt}{\mbox{$\D t_t$}}

\newcommand{\orb}{{\tt orbitN}}
\newcommand{\snv}{{\tt snvec}}
\newcommand{\zenurl}{\url{doi.org/10.5281/zenodo.8021040}}
\newcommand{\giturl}{\url{github.com/rezeebe/orbitN}}
\newcommand{\myurl}{\url{www2.hawaii.edu/~zeebe/Astro.html}}
\newcommand{\npurl}{\url{www.ncei.noaa.gov/access/paleo-search/study/39199}}
\newcommand{\npturl}{\url{www.ncdc.noaa.gov/paleo/study/35174}}
\newcommand{\snvurl}{\url{github.com/rezeebe/snvec}}
\newcommand{\walturl}{\url{http://nm2.rhul.ac.uk/Milankovitch1.html}}

\ifTWO
\def\bls{0.7}
\def\fls{0.7}
\else
\def\bls{1.7}
\def\fls{1.2}
\fi
\renewcommand{\baselinestretch}{\bls}
\def\otp{\sm{40}\%}

\begin{document}

\title{\bf Milankovi{\'c} Forcing in Deep Time}

\authors{Richard E. Zeebe$^{1,*}$ and Margriet L. Lantink$^2$}
     \normalfont \noi
     $^*$Corresponding Author.\\ \noi {\small
     $^1$School of Ocean and Earth Science and Technology, 
     University of Hawaii at Manoa, 
     1000 Pope Road, MSB 629, Honolulu, HI 96822, USA. 
     zeebe@soest.hawaii.edu      \\[2ex] \noi
     $^2$University of Wisconsin - Madison, Department of Geoscience,
     1215 W. Dayton St., Madison, WI 53706, USA.
     lantink@wisc.edu}  \\[2ex] \noi
     \today \\[2ex] \noi
     Final revised version. 
     In press, {\it Paleoceanography and Paleoclimatology} \\[3ex]

\vspace*{20ex}
\renewcommand{\baselinestretch}{1.0} \fns
\noi
{\bf Key Points}
\begin{itemize}
\item
We provide orbital eccentricity, inclination, oliquity, and climatic precession for use in paleostudies/climate models over the past 3.5 Gyr
\item
The long eccentricity cycle (previously used as "metronome") can become unstable on long time scales
\item
Earth's past obliquity forcing/amplitude was significantly reduced. We predict reduced obliquity power with age in stratigraphic records
\end{itemize}
\renewcommand{\baselinestretch}{1.0} \fns
\noi
{\bf Keywords}\\[-1ex]

Milanković Theory (4969), Astronomical Forcing (4910), Paleoclimatology and Paleoceanography (0473), Sedimentary Geochronology (1165), Astrochronology, Cyclostratigraphy, Solar System

\renewcommand{\baselinestretch}{1.2} \normalsize 
       
\begin{abstract}
Astronomical (or Milankovi{\'c}) forcing of the Earth system is key to 
understanding rhythmic climate change on time~scales $\gsim$10$^4$~y.
Paleoceanographic and paleoclimatological applications concerned with 
past astronomical forcing rely on astronomical calculations (solutions), 
which represent the backbone of cyclostratigraphy and astrochronology.
Here we present state-of-the-art
astronomical solutions over the past 3.5~Gyr. Our goal is 
to provide tuning targets and templates for interpreting deep-time 
cyclostratigraphic records and designing external forcing functions in 
climate models. Our approach yields internally consistent orbital and 
precession-tilt solutions, including fundamental solar 
system frequencies, orbital eccentricity and inclination, lunar distance, 
luni-solar precession rate, Earth's obliquity, and climatic precession.
Contrary to expectations, we find that the long eccentricity cycle 
(previously assumed stable and labeled ``metronome'', recent 
period \sm{405}~kyr), can become unstable on long time~scales. Our results 
reveal episodes during which the long eccentricity cycle is very 
weak or absent and Earth's orbital eccentricity and climate-forcing 
spectrum are unrecognizable compared to the recent past. For the ratio 
of eccentricity-to-inclination amplitude modulation (frequently 
observable in paleorecords) we find a wide distribution around the recent 
2:1 ratio, i.e., the system is not restricted to a 2:1 or 1:1 resonance 
state. Our computations show that Earth's obliquity was lower and 
its amplitude (variation around the mean) significantly reduced in 
the past. We therefore predict weaker climate~forcing at obliquity 
frequencies in deep time and a trend toward reduced obliquity power 
with age in stratigraphic records. For deep-time stratigraphic and 
modeling applications, the orbital parameters of our 3.5-Gyr 
integrations are made available at 400-year resolution.
\end{abstract}

\renewcommand{\baselinestretch}{\bls}

\ifTWO \twocolumn \small \fi

%
%
%
\section{Introduction \label{sec:intro}}

In 1941, Milankovi{\'c} commented on the motivation for his work
on insolation: 
``If it were actually possible [$\ldots$] to create a 
mathematical theory by means of which one could track the effect of 
insolation in space and time, one would be able to 
determine the most important basic features of the Earth's climate
computationally.'' \citep[][see~\ref{sec:ger}]{milank41}.
Today, the so-called Milankovi{\'c} cycles are known as periodic changes 
in Earth's orbital parameters, causing rhythmic climate change on Earth
on time scales $\gsim$10$^4$~y.
Milankovi{\'c}\,'s primary aim was to apply his astronomical theory
to the ice age problem. Since then, numerous studies have investigated 
the astronomical forcing of climate on \tcr{time scales ranging
from half-precession periods to Gyrs}, primarily relying on astronomical 
calculations/solutions, which today represent the backbone of 
astrochronology and cyclostratigraphy \tcr{\citep[for recent summaries, 
see][]{montenari18,hinnov18,devleesch24}}.
While undoubtedly a visionary, Milankovi{\'c} may not have foreseen
using digital computer clusters to numerically determine astronomical 
forcing of Earth's climate over billions of years of its history
(the undertaking of the present study).

Cyclostratigraphy, the study of astronomically induced cycle patterns 
expressed in stratigraphic sequences, enables reconstructing
the Earth System's intricate response to Milankovi{\'c} 
forcing and provides a tool for establishing high-resolution 
temporal frameworks for ordering/dating 
geologic history (astrochronologies). Beyond \sm{50}~Ma, 
astrochronology may be used to reconstruct the chaotic evolution 
of orbital cycles such as the unstable, very long eccentricity cycle
\citep[\sm{2.4}~Myr at present, e.g.,][]{olsen19,mameyers17,zeebelourens19}
and in some cases individual fundamental solar system frequencies 
\citep[e.g.][]{meyers18,olsen19}.
Moreover, the cyclostratigraphic record of changing precession and 
obliquity frequencies permits reconstructing the long-term 
(tidal) evolution of the Earth-Moon system. The latter approach is 
especially valuable in Precambrian records, when changes in precession 
periods and their ratios to eccentricity periods are more pronounced
\citep[e.g.,][]{zhang15orb,meyers18,lantink19,lantink22,lantink24}.
However,
cyclostratigraphic reconstructions and astrochronology purely
based on observational data in deep time
is challenging in the absence of astronomical tuning targets 
due to the uncertain long-term tidal evolution of the 
Earth-Moon system and the solar system's chaotic nature
\citep{berger89ast,laskar04NatB,zeebe17aj}. 
Conversely, observational studies generate fundamental knowledge
and several hard data constraints on the Earth-Moon and solar
system history, which in turn, help \tcr{to constrain} the 
astronomical solutions \citep[e.g.][]{zeebelourens19,lantink22}.
Thus, cyclostratigraphic data and astronomical solutions 
are inherently valuable and in high demand, particularly in deep 
time.

In this study, we present astronomical solutions from 
state-of-the-art solar system integrations 
over the past 3.5~Gyr within a single, internally consistent
framework.
We highlight important features of the overall nature of, and 
the dominant frequency components involved in, deep-time
Milankovi{\'c} forcing and discuss which cycles may be (un)suitable
for developing deep-time astrochronologies. Our astronomical
solutions provide examples of possible characteristic forcing 
patterns to assist in the interpretation of deep-time 
cyclostratigraphic records and the design of external forcing 
functions in climate models.
We supply two types of internally consistent astronomical solutions, 
orbital solutions (OSs) and precession-tilt (PT) solutions (see 
Sections~\ref{sec:os} and~\ref{sec:pts}). OS dynamics have important
effects on (and are a prerequisite for) PT solutions (the 
opposite effect is minor, see Section~\ref{sec:tfpast}). For example, 
amplitude variations in Earth's 
orbital inclination (due to OS dynamics) are reflected in obliquity ---
most evidently during intervals of reduced amplitude variation 
\citep[see][]{zeebe22aj}. 
Similarly, amplitude variations in eccentricity (due to OS 
dynamics) are reflected in climatic precession.
Our combined OS and PT solutions and their analyses yield diagnostic 
features of deep-time orbital forcing parameters that are also 
key in more recent cyclostratigraphic and astrochronologic practices. 
The diagnostics include the different eccentricity cycles (ECs) and
the relative periodicities and amplitudes of P: T: SEC: LEC: VLEC
(Precession: Tilt: Short~EC: Long~EC: Very~Long~EC), as well as the 
stability of the LEC (previously assumed stable and used as a ``metronome'' 
in astrochronology).

\subsection{Specific goals, benefits, and outcome provided}

Our main goal is to investigate deep-time Milankovi{\'c} 
forcing and make our results available
to the community. Our findings are based on long-term 
ensemble integrations ($N = 64$, see below), exploring the possible 
solution/phase space of the solar system. Hence our study
provides characteristic features of long-term Milankovi{\'c} 
forcing and 64 individual solutions, not a single 
forcing function (as prohibited by dynamical chaos).
From our OS and PT solutions, we supply outcome including
the fundamental (secular) solar system frequencies
($g_i$ and $s_i$),
lunar distance ($a_L$),
luni-solar precession rate ($\Psi$),
obliquity (\obl),
and climatic precession (\pbar)
(see Table~\ref{tab:notval}).
Importantly, we integrate the equations of motion for Earth's
spin axis over 3.5~Gyr, yielding full solutions for \obl\
and \pbar\ (see Section~\ref{sec:pts}).
We also include error estimates and a comparison to 
\citet{waltham15}'s results for $a_L$, $\Psi$, and
averaged \obl\ (see Figs.~\ref{fig:PT} and~\ref{fig:walt}).
The results of our 3.5-Gyr OS and PT integrations 
($N = 64$), including eccentricity, inclination, obliquity, and 
climatic precession are made available at 400-year 
resolution at \npurl\ and \myurl.

\def\cau{$1.495978707\e{11}$}
\def\cgm{$1.32712440041\e{20}$}
\def\ugm{$\rm m^3~s^{-2}$}
\def\com{$7.292115\e{-5}$}
\def\crn{$3.8440\e{8}$}
\def\cen{$0.00327381$}
\def\cse{$328900.5596$}
\def\cel{$81.300568$}
\def\cgk{$0.9925194$}
\def\psn{$50.384815$}
\def\psnM{$7.597$}
\def\gdp{$-0.0192$}
\def\me{$5.9720\e{24}$}
\def\re{$6378.136\e{3}$}
\def\ks{$0.942$}
\def\kt{$0.3$}

\begin{table*}[p]
\renewcommand{\baselinestretch}{0.85} \small
\caption{Notation and values used in this paper. \label{tab:notval}}
\begin{tabular}{lllll}
\hline
Symbol       & Meaning              & Value I/A & Unit  & Note  \\
\hline     
\obl         & Obliquity angle          &      &     &       \\
\obl$_0$     & Obliquity Earth $t_0$    & 23.4392911 & deg & \citet{fraenz02} \\
\prc         & Precession angle         &      &       &       \\ 
\alp         & Precession constant      &      &       & see \ref{sec:kbet}   \\
\vs          & Spin vector              &      &       & unit vector          \\
\vn          & Orbit normal Earth       &      &       & unit vector          \\
\vb          & Orbit normal Lunar       &      &       & unit vector          \\
$i_L$        & Inclination lunar orbit  &      &  
                                        & $\cos i_L =   (\vb \* \vn) = y$     \\
\epss        & Mutual obliquity         &      &  
                                        & $\cos \epss = (\vs \* \vb) = z$     \\
$e$          & Orbital eccentricity     &      &       &       \\ 
$\vpi$       & Orbit LP $^a$            &      &       &       \\ 
\ombar       & Orbit LPX $^b$           &      &       &       \\ 
$I$          & Orbital inclination      &      &       &       \\ 
\Om          & Orbit LAN $^c$           &      &       &       \\ 
\pbar        & Climatic precession      &      &       & $\pbar = e \sin \ombar$ \\ 
\ggp         & Geodetic precession      & \gdp & \asy  & \citet{capitaine03} \\ 
$g_i$, $s_i$ & Secular frequencies      &      & \asy  &       \\ 
au           & Astronomical unit        & \cau & m     &       \\ 
$GM$         & Sun GP $^d$              & \cgm & \ugm  &       \\ 
$M/(m_E+m_L)$& Mass ratio $^e$          & \cse & --    &       \\
$m_E/m_L$    & Mass ratio $^e$          & \cel & --    &       \\
$\mu$        & Lunar GP $^d$            &      &       & $\mu = GM(m_E+m_L)/M$ \\
\vL          & Angular Momentum         &      &       &       \\ 
\vT          & Torque                   &      &       &       \\ 
\om          & Earth's angular speed    & \com & s\pmo & at $t_0$ \\
$a_L$        & Earth-Moon DP $^f$       &      & m     &       \\ 
$a_{L0}$     & Earth-Moon DP $^f$ $t_0$ & \crn & m     & \citet{quinn91}\\ 
$n_L$        & Lunar mean motion        &      &       & $n_L = (\mu/a_L^3)^{1/2}$ \\ 
$A,C$        & Moments of inertia $^g$  &      &       &       \\
$E_{d0} =(C-A)/C$     
             & Earth's dyn. ellipticity & \cen & --    & at $t_0$, see text  \\ 
$g_L$        & Lunar orbit factor       & \cgk & --    & see text            \\
$\Psin = \dot{\phi}_0$      
             & Luni-solar prec. $t_0$   & \psn & \asy  & \citet{capitaine03} \\
$m_E$        & Mass Earth               & \me  & kg    &       \\
$R_E$        & Radius Earth             & \re  & m     &       \\
$k_s$        & Love No.\ Secular        & \ks  & --    & $^h$  \\
\ktt         & Love No.\ TE $^k$        & \kt  & --    & $^h$  \\
\dtt         & Tidal time lag           &      &       & \citet{mignard81}   \\
\del         & Tidal phase lag          &      &       & $^h$  \\
$K$          & Solar factor for \alp\ $^l$ &   &       & see \ref{sec:kbet}  \\
\bet         & Lunar factor for \alp\ $^l$ &   &       & see \ref{sec:kbet}  \\
\hline 
\end{tabular}

{\fns
$^a$ LP  = Longitude of Perihelion.
$^b$ LPX = LP from the moving equinox.
$^c$ LAN = Longitude of Ascending Node.
$^d$ GP = Gravitational Parameter.
$^e$ Index $E$ = Earth, $L$ = Lunar.
$^f$ DP = Distance Parameter.
$^g$ Earth's equatorial ($A$) and polar ($C$) moment of inertia.
$^h$ For details, see \citet{macdonald64,goldreich66,lambeck80,baenas19}.
$^k$ TE = tidal-effective.
$^l$ Factors for \alp\ related to solar and lunar torque (see \ref{sec:kbet}).
}
\end{table*}

\renewcommand{\baselinestretch}{\bls}\normalsize

\ifTWO \small \fi

The present astronomical solutions are designed for
\tcr{deep-time applications (order $t \ \lsim$$-10^8$~y)}. For 
the past 100~Myr (300~Myr), we recommend the 
orbital solutions ZB$18$a 
and ZB$20x$, which have been constrained by geological data and
are more accurate on that time scale \citep{zeebelourens19,
zeebelourens22epsl}. 
ZB$18$a and ZB$20x$ are not incompatible with the
current results but feature slightly different attributes,
including asteroids and a different timestep.
Precession-tilt solutions for the
past 100~Myr are available at \myurl\ and \npturl\ and can 
be generated/ customized using \snv\ (\snvurl) 
\citep{zeebelourens22pa}.

%
%
%
\section{Orbital Solutions \label{sec:os}}

For the orbital solutions,
we performed state-of-the-art solar system integrations,
including the eight planets and Pluto, a lunar contribution, 
general relativity, the solar quadrupole moment, and solar mass loss
\citep{zeebe17aj,zeebelourens19,zeebe22aj,zeebe23aj,
zeebelantink24aj}.
Initial conditions at time $t_0$ were taken from the
latest JPL ephemeris DE441 \citep{park21de} and the equations
of motion were numerically integrated to $t = -3.5$~Gyr
(beyond $-3.5$~Gyr very few geologic records are available and
parameters such as the lunar 
distance have large uncertainties, see Section~\ref{sec:tfpast}).
Owing to solar system chaos, the solutions diverge around $t = -50$~Myr,
which prevents identifying a unique solution on time scales 
$\gsim$10$^8$~y, see \citep{laskar04NatB,zeebe17aj,
zeebelourens19,zeebelourens22epsl}. Hence we present
results from long-term ensemble integrations to explore the 
possible solution/phase space of the system.
Importantly, because of the chaos, each \sm{$10^8$}~y interval 
of the integrations represents a snapshot of the system's 
general/possible
behavior that is largely independent of the actual numerical time of 
a particular solution (provided here that $t < -\tauot$, where 
\tauot\ is of order $10^8$ to $10^9$~y, see below).
In other words, a numerical solution's behavior around,
say, $t = -1.5$~Gyr may represent the actual solar system 
around $t = -600$~Myr and so on. 
Our approach allows for full solar system dynamics and full
dynamical chaos and is fundamentally different from 
\citet{hoang21} who used a simplified, secular model 
(truncated at 2nd order in masses and 5th order in 
eccentricities/inclinations).

\subsection{Numerical integrator and physical setup}

Solar system integrations were carried out following
our earlier work 
\citep{zeebe15apjA,zeebe15apjB,zeebe17aj,zeebelourens19,zeebe22aj} 
with our integrator package {\tt orbitN (v1.0)} \citep{zeebe23aj}.
The open source code is available at \zenurl\ and \giturl.
The methods, physical setup, and our integrator package 
used here have been extensively tested and compared against other 
studies \citep{zeebe17aj,zeebelourens19,zeebe22aj,
zeebe23aj}; for more information, see~\ref{sec:num}.

\subsection{Ensemble integrations}

We performed ensemble integrations of the solar system
with a total of $N = 64$ members. Note that a larger $N$ is 
not necessarily advantageous  for the current application. 
For one, our goal is to explore the system's general behavior
and provide an abundance of possible values for the fundamental 
solar system frequencies (secular $g$- and $s$-frequencies).
For instance, time series analysis
of consecutive 20-Myr intervals of our simulations (see below)
provide 64 $\*$ (3,500/20) = 11,200 values for
each frequency, which is plenty. The analysis of some
frequencies is cumbersome and requires individual inspection
and manual work (see below), which becomes unfeasible for large 
$N$. Also, regarding the occurrence of, e.g., \sigot\ resonances 
(see~\ref{sec:lec} and \citet{zeebelantink24aj}), 
our ensemble integrations sample a frequently occurring
phenomenon (\otp\ of solutions), not a rare event
\tcr{such as the destabilization of Mercury's orbit},
which would require large $N$ \citep[e.g.,][]{laskar09,
zeebe15apjB,abbot23}.
\tcr{
The fact that the \sigot\ resonance was (1) recognized previously 
\citep{lithwick11,mogavero22} --- although not its effect on 
the LEC \citep[see][]{zeebelantink24aj} --- and (2) found in \otp\ of 
our solutions suggests that entering the \sigot\ resonance on
long time scales is a common (not rare) dynamical feature.
}
Different solutions were obtained by offsetting
Earth's initial position by a small distance 
(largest offset $\D x_0 \simeq 1\x 10^{-12}$~au), which is 
within observational uncertainties \citep{zeebe15apjB,
zeebe17aj}. The different $\D x_0$ lead to complete
randomization of solutions on a time scale of \sm{50}~Myr
due to solar system chaos. We also tested different 
histories of the Earth-Moon distance ($a_L$), which has 
little effect on our OS results (see Section~\ref{sec:tfpast}). 
Because of scarce geological records and
the large uncertainties in $a_L$ prior to 
\sm{$3.5$}~Ga, we restrict our integrations to $t = -3.5$~Gyr.
Our solutions are available at \npurl\ and \myurl.

\subsection{Past Earth-Moon distance}

Our solar system integrations
include a lunar contribution, i.e., a gravitational quadrupole 
model of the Earth-Moon system \citep{quinn91,varadi03,zeebe17aj,
zeebe23aj}. The lunar contribution has a relatively small effect 
on the overall orbital dynamics of the solar system, yet the 
integration requires the Earth-Moon distance ($a_L$) as a parameter 
at a given time in the past (details and technical aspects are 
discussed in Section~\ref{sec:tfpast}).

\subsection{Time series analysis of astronomical solutions}

The solar system's fundamental $g$- and $s$-frequencies
(aka secular frequencies)
were determined from the output of our numerical integrations
using fast Fourier transform (FFT) over consecutive 20-Myr 
intervals. For the spectral analysis, we used Earth's 
orbital elements and the classic variables:
\beqn
h =  e \sin(\vpi)         \quad & ; & \quad
k =  e \cos(\vpi)         \label{eqn:hk} \\
p = \sin (I/2) \ \sin \Om \quad & ; & \quad
q = \sin (I/2) \ \cos \Om \label{eqn:pq} \ ,
\eeqn
where $e$, $I$, $\vpi$, and $\Om$ are eccentricity, 
inclination, longitude of perihelion, and longitude
of ascending node, respectively. The variables
$h,k,p,q$ are useful analytically and obey the
relationships $h^2 + k^2 = e^2$ and $p^2 + q^2 = 
\sin^2 (I/2)$ \citep[e.g.,][]{murraydermott99}.
The spectra for Earth's $h,k$ and $p,q$, for example, show
strong peaks at nearly all $g$- and $s$-frequencies,
respectively \citep[see][]{zeebelantink24aj}, i.e., analyzing
only Earth's spectrum may suffice (instead of all 
planetary spectra), depending on the application.
The $g$- and $s$-modes are loosely related 
to the apsidal and nodal precession of the planetary orbits 
\citep[see][]{zeebelantink24aj}.
However, there is generally no simple one-to-one relationship 
between a single mode and a single planet, 
particularly for the inner planets. The system's motion is 
a superposition of all modes, although for the outer planets, 
some modes are dominated by a single planet. 

%
%
%
\section{Precession-Tilt Solutions and Framework \label{sec:pts}}

\subsection{Theoretical PT framework and previous studies}

The current approach provides combined OS and PT solutions, the 
latter of which are obtained within a theoretical framework
\tcr{(Framework~I hereafter)}
that has been applied to the Earth-Moon system, Pluto-Charon,
exoplanets, etc. \citep[e.g.,][]{macdonald64,goldreich66,mignard81,
touma94,atobe07,chengpeale14,downey23}. The emphasis is on long-term 
physical solutions for
the planetary spin axis, the satellite's orbital inclination,
and (here) explicit obliquity and precession
solutions. In a number of related, but somewhat distinct studies,
the emphasis has shifted toward a detailed model framework for
the Earth-Moon's tidal evolution \tcr{(Framework~II hereafter)}, 
particularly the history of
ocean tidal dissipation \citep[e.g.,][]{webb82,hansen82,kagan94,
motoyama20,daher21,tyler21,farhat22}. The advance in 
modeling past ocean tidal evolution provided by these studies 
is clearly
desirable. Importantly, however, studies of both frameworks 
(regardless of emphasis) ultimately rely on fitting results to external 
parameters and observational data in order to provide realistic astronomical
parameters such as lunar distance, precession frequency, etc. on 
Gyr-time scale. For instance, \citet{daher21} explore 
tidal energy dissipation rates based on reconstructed basin 
paleogeometries but are unable to collapse the lunar distance
to near zero at \sm{4.4}~Ga (lunar age).
\citet{tyler21} fits two parameters
(effective ocean depth and dissipation time scale) to observational 
data including length of day/month and lunar distance.
\citet{farhat22} also fit two parameters (uniform effective ocean 
depth and effective dissipation frequency) to the lunar 
age and the present lunar recession rate. In addition,
\citet{farhat22} assume a smooth transition \tcr{from} a global
ocean planet prior to 1~Ga. Unfortunately, observational data 
to verify or falsify the different tidal evolution models are 
extremely sparse, especially prior to \sm{1}~Ga. If restricted 
to robust data sets from cyclostratigraphic studies, only two 
data points are available with ages older than 1~Ga (see 
Fig.~\ref{fig:rr0}b). 
\tcr{For more information on the cyclostratigraphic 
studies \citep{meyers18,lantink22,soerensen20,devleesch23L},
see Section~\ref{sec:tfpast}.
}

As detailed below, the current approach to obtain PT solutions 
follows Framework~I,
using an internally consistent physical model with a single fit 
parameter, the tidal time lag \dtt\ \citep{mignard81,
touma94}. Analog to Framework~II, \dtt\ is fit to match
theoretical results with observational data (here cyclostratigraphic 
data for the lunar distance, see Fig.~\ref{fig:rr0}). At this time, 
no observational evidence is available to decide whether 
Framework~I or~II provides more accurate/realistic results for 
the current application over the full past 3.5~Gyr. For the 
present theoretical approach (PT solutions), Framework~I is much 
preferred because the necessary assumptions and 
mathematical approach can be stated clearly and succinctly
(see Section~\ref{sec:pts} and~\ref{sec:trqs}). 
If desired, results of tidal evolution models 
may be evaluated within the current framework 
(see~\ref{sec:tiderr}, Fig.~\ref{fig:PTF22}).

\subsection{Precession-Tilt Solutions}

The precession-tilt (PT) solutions computed here are based on
our earlier work \citep{zeebelourens22pa,zeebe22aj} using
the \snv\ code, see \snvurl\ and~\ref{sec:kbet}. 
The original \snv\ code was
developed to provide PT solutions over the past 100~Myr or so,
following \citet{quinn91}.
While the code includes parameters for tidal dissipation and 
dynamical ellipticity, it lacks proper dynamical equations 
for the long-term evolution of the Earth-Moon system 
(including non-linear evolution of lunar distance, Earth's
spin, luni-solar precession, etc.). Hence
we added several features and differential equations to
provide PT solutions on Gyr-time scale.
The integration of the long-term precession equations
follows the classical work of 
\citet{macdonald64,goldreich66,mignard81}, shown to capture
the essential dynamics \citep[e.g.,][]{touma94} and
appropriate for a variety of applications \citep[e.g.,][]
{atobe07,chengpeale14,downey23}, as well as the current 
one (see below).

\subsection{Differential equations for $L$ and $a_L$ 
\label{sec:deqs}}

We added differential equations to the \snv\ code
for the magnitude of Earth's angular momentum 
$L$ and the Earth-Moon (lunar) distance $a_L$. For variables
and symbols, see Table~\ref{tab:notval}. The equation for $L$ 
reads \citep[see Section~\ref{sec:tfeq} and][]{goldreich66}:
\beqn
\dot{L} = \vT_E \* \vs & = & T_2 \sin \obl^* + T_3 \cos \obl^* \ ,
\eeqn
where the mutual obliquity $\obl^*$ is the angle between Earth's
spin (unit) vector \vs\ and the lunar orbit normal \vb;
$T_2$ and $T_3$ are torque components 
\tcr{(see Section~\ref{sec:tfeq} and~\ref{sec:trqs})}.
The current application does not require resolving
the lunar orbit precession (\sm{18.6}~y). Hence for the
time-averaged lunar orbit normal, we may take
$\bbar \simeq \vn$, where \vn\ is Earth's orbit 
normal (in that approximation, $\overline{\obl^*} \simeq 
\obl$). The equation for $a_L$ reads \citep{goldreich66}:
\beqn
\dot{a}_L =  2 \ a_L \v{T}_L \* \vb / l
          =  2 \ a_L T_{3L} / (m_L \sqrt{\mu a_L}) \ ,
\label{eqn:dotaL}
\eeqn
where $T_{3L} = -T_3$
\tcr{(see Section~\ref{sec:tfeq} and~\ref{sec:trqs})}, 
$l = m_L \sqrt{\mu a_L}$ 
is the lunar angular momentum and $\mu = GM \* (m_E+m_L)/M$.
Integration of the above differential equations yield 
$a_L(t)$ and $\om(t) = L(t)/C(t)$, from which $K$, \bet,
\alp, and $\Psi$ can be calculated (see Eq.~(\ref{eqn:psi})
and~\ref{sec:kbet}).
For error estimates and a comparison to \citet{waltham15}'s 
results for $a_L$, $\Psi$, and averaged \obl, see 
Figs.~\ref{fig:PT} and~\ref{fig:walt}.

\subsection{Moment of Inertia}

Earth's angular spin, \om, is calculated from $L = C \om$, where 
$C$ is the moment of inertia \citep{macdonald64,goldreich66}:
\beqn
C = I\ \{ 1 +  (2 k_s R_E^5/9GI) \ \om^2 \}
\label{eqn:CIN}
\eeqn
and $I =  0.33 \ m_E \ R^2_E$. Note that $C$ depends on \om,
yet changes in \om\ and $C$ are small per integration step.
Thus, numerically it suffices to update $\om = L/C$ and 
Eq.~(\ref{eqn:CIN}) each time the derivative routine is 
called, which occurs multiple times per integration step.

\subsection{Tidal friction equations \label{sec:tfeq}}

The tidal friction equations in, e.g., \citet{goldreich66} 
and \citet{touma94} are given in terms of
obliquity/inclination angles. However, our \snv\ code 
integrates differential equations for Earth's spin
vector \vs\ \citep[see][]{quinn91,zeebelourens22pa,zeebe22aj}.
In the following, we therefore derive a tidal friction equation
for \vs. Let's write Earth's angular momentum vector as:
\beqn
\v{L} = L \v{s} \ ,
\eeqn
where (as mentioned above) \vs\ is a unit vector. It follows
for the torque ($d\v{L}/dt = \v{T}_E$):
\beqn
d/dt (L \v{s}) = \dot{L} \v{s} + L \dot{\v{s}} = \v{T}_E \ .
\label{eqn:dotLs}
\eeqn
Now dot Eq.~(\ref{eqn:dotLs}) by $\v{s}$ and note that
$\vs \* \vs = 1$. Also, for \vs\ to remain a unit vector, 
any change must be perpendicular, i.e., $\dot{\vs} \* \vs = 0$
and hence:
\beqn
\dot{L} = \v{T}_E \* \v{s} \ .
\eeqn
Inserting into Eq.~(\ref{eqn:dotLs}) then yields an equation for 
$\dot{\vs}$, as desired:
\beqn
\dot{\vs} = [ \v{T}_E - (\v{T}_E \* \vs) \ \vs ] / L \ .
\eeqn
The $\vT_E$ terms are given by ($T_1 = 0$,
\tcr{see~\ref{sec:trqs}} for $T_2$ and $T_3$):
\beqn
\vT_E & = & T_1 \ \v{e}_1 + T_2 \ \v{e}_2 + T_3 \ \v{e}_3 \\
\vT_E \* \vs & = & T_2 \ \sin \obl^* + T_3 \ \cos \obl^* \ ,
\eeqn
where the $\v{e}_i$ are unit vectors forming a coordinate system 
in which the torque components are conveniently expressed 
\citep{goldreich66}. The $\v{e}_i$ are given by ($\v{e}'_i$
are used to project $T'_i$, see~\ref{sec:trqs}):
\ifTWO
\beqn
\v{e}_1  = (\vs\x \vb)            / \sin \epss ;  && 
\v{e}_2  = (\vs - \vb \cos \epss) / \sin \epss ; \nonumber \\ &&
\v{e}_3  =  \vb \ \\
\v{e}'_1 = (\vs\x \vn)            / \sin \eps ; &&
\v{e}'_2 = (\vs - \vn \cos \eps ) / \sin \eps ; \nonumber \\ &&
\v{e}'_3 =  \vn \ .
\eeqn
\else
\beqn
\v{e}_1  = (\vs\x \vb)            / \sin \epss \quad ; \quad
\v{e}_2  = (\vs - \vb \cos \epss) / \sin \epss \quad ; \quad
\v{e}_3  =  \vb \\
\v{e}'_1 = (\vs\x \vn)            / \sin \eps  \quad ; \quad
\v{e}'_2 = (\vs - \vn \cos \eps ) / \sin \eps  \quad ; \quad
\v{e}'_3 =  \vn \ .
\eeqn
\fi

\subsection{Tidal torques \label{sec:mit}}

Within the present framework, it is possible to use different tidal
torque expressions \citep[e.g.,][]{macdonald64,goldreich66,mignard81,
touma94}. For the most part, the differences obtained for 
the different torque expressions are relatively small 
for $a_L \ \gsim \ 30 R_E$, except for Earth's obliquity. 
Over this range (which is explored here), uncertainties in 
the tidal friction evolution (if unconstrained by data) are 
usually larger than those arising from the torque expressions.
We have
tested two sets of torques (Mignard's and MacDonald's torques,
see~\ref{sec:trqs} for equations) and implemented both in the 
\snv\ code. MacDonald's torques are relatively easy to implement 
but do not include solar tides and solar-lunar cross terms
\citep[see][]{touma94}. Also, some fundamental issues with
MacDonald's approach have been noted \citep[e.g.][]{efroimsky13}.
More importantly, the solar-lunar cross terms (not included in
MacDonald's torques) have a significant effect on the obliquity 
evolution and tend to align the spin axis with the orbit normal,
which is relevant here (see Fig.~\ref{fig:PT}).
Hence we selected 
Mignard's tidal torques, including cross terms, as our default 
option \citep[see][and~\ref{sec:trqs}]{mignard81,touma94}.

\subsection{Tidal friction in the past \label{sec:tfpast}}

\begin{figure}[t]
\renewcommand{\baselinestretch}{\fls}
\ifTWO
\vspace*{-36ex} \hspace*{-10ex} \def\vs{-32ex}
\else
\vspace*{-50ex} \hspace*{-05ex} \def\vs{-43ex}
\fi
\includegraphics[scale=\f1scale]{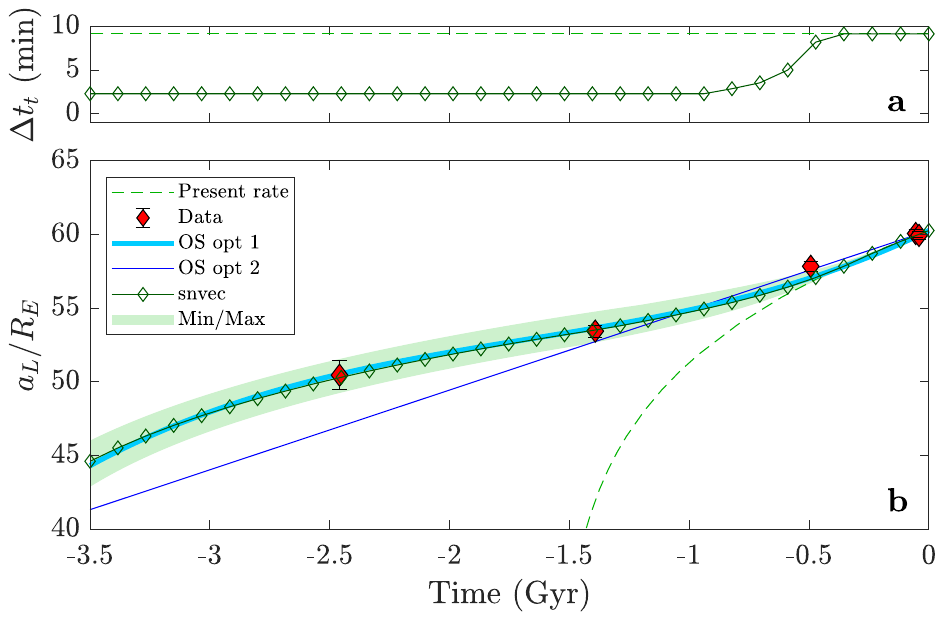}
\vspace*{\vs}
\caption{\scs
(a) Tidal time lag \dtt\ (diamonds) used in Mignard's torques
including solar-lunar cross terms 
(see~\ref{sec:trqs}). Dashed line shows constant (present) \dtt.
(b) Past Earth-Moon distance ($a_L$) in units of Earth radii ($R_E$).
Red diamonds: Observational estimates 
based on robust data sets from cyclostratigraphic studies 
\tcr{(see Section~\ref{sec:tfpast})}.
Cyan and blue lines: \tcr{Options~1 and~2} used in the orbital 
solution (OS). Blue:
linear extrapolation into the past starting with $\dot{a}_L$
close to the present rate. Cyan: 3rd-order polynomial fit 
to observations (see Eq.~(\ref{eqn:poly})). 
Using $a_L$ based on the blue and cyan lines
made essentially no difference in our OSs 
\tcr{(see text Section~\ref{sec:tfpast})}.
Green lines/symbols: Integration of precession equations with \snv.
Light green area (Min/Max): Error envelope for $a_L$ reflecting
cyclostratigraphic data errors, see Section~\ref{sec:pts} and 
Fig.~\ref{fig:walt}.
Green dashed: starting at present rate $\dot{a}_{L0} = 3.82$~cm~y\pmo\ with
constant tidal time lag ($\dtt \simeq 9$~min) in the past (see (a)), yields
the (well-known) unrealistic past $a_L$. Dark green diamonds: using 
variable $\dtt$ in the past from (a), internally consistent with 
OS option~1, see text.
\label{fig:rr0}
}
\end{figure}

Before discussing tidal friction in the past, we first explain 
a technical aspect of our approach.
The PT solution provides the lunar distance, $a_L$, at a given time 
in the past (see Eq.~(\ref{eqn:dotaL})), based on which precession 
and obliquity are computed. Now solving the PT equations requires
output from the OS as input, while
the OS requires $a_L$ as input for the gravitational 
quadrupole model of the Earth-Moon system \citep{quinn91,varadi03,
zeebe17aj,zeebe23aj}. Importantly, however, the lunar contribution 
has a relatively small effect on the overall OS dynamics. 
Furthermore, in terms of 
computations and data handling, it is much more convenient to run 
OS and PT integrations separately. Thus, we prescribed $a_L$ as input
for the OS (including different options, see below) and performed 
ensemble integrations for the orbital part first. Next,
we used the OS output as input for the PT integrations, 
which allowed for easy parameter variations.
For the prescribed $a_L$ (OS input), we included an option based on
observational data (see Fig.~\ref{fig:rr0}) that is internally 
consistent between OS and PT solution.

Integration of the classic precession equations with \snv\ 
(see Sections~\ref{sec:deqs}-\ref{sec:mit}) and a constant (present) 
tidal time lag of $\dtt \simeq 9$~min ($\dot{a}_{L0} = 3.82$~cm~y\pmo)
leads to the well-known problem of 
unrealistically small $a_L$ in the past (Fig.~\ref{fig:rr0}, green
dashed line). Thus, for the PT integrations, we used a variable
\dtt\ in the past (Fig.~\ref{fig:rr0}a) that yields $a_L$ consistent 
with both observational data and the
prescribed $a_L$ input option~1 for the OS (Fig.~\ref{fig:rr0}b).
(For an alternative \dtt\ history and evaluation of tidal evolution 
models, see Fig.~\ref{fig:PTF22}.)
Note that due to its small effect, selecting OS input 
option~1 or~2 for $a_L$ made essentially no difference in our long-term 
OS ensembles.
For the observational constraints, we selected robust data sets based 
on the reconstruction of Earth's axial precession frequency obtained 
by cyclostratigraphic studies \citep{meyers18,lantink22,soerensen20,
devleesch23L} (Fig.~\ref{fig:rr0}b).
\tcr{
Other methods for estimating past $a_L$ (and precession frequency)
include the analysis of tidal 
rhythmites and fossil growth laminae, but these approaches are 
generally associated with large uncertainties and ambiguities in 
interpretation, especially for Precambrian time intervals 
\citep{lantink22,laskar24}. 
Our selection of cyclostratigraphic $a_L$ estimates was based 
on two main quality criteria
\citep[for further discussion, see][]{sinnesael19}. 
(1) The presence of clearly developed, visually identifiable 
rhythms in studied proxy records, exhibiting expected Milankovi{\'c} 
cycle (period) ratios and amplitude modulation relationships, 
supported by results of time-series analysis and statistical 
hypothesis testing (i.e., records with a high signal-to-noise
ratio). (2) Consideration of additional chronostratigraphic data 
(e.g., radioisotopic ages, magnetostratigraphy, biostratigraphy) 
that independently support Milankovi{\'c} interpretations of 
observed stratigraphic patterns.}

\begin{figure*}[p]
\vspace*{-50ex} \hspace*{-45ex}
\mbox{
\includegraphics[scale=1.1]{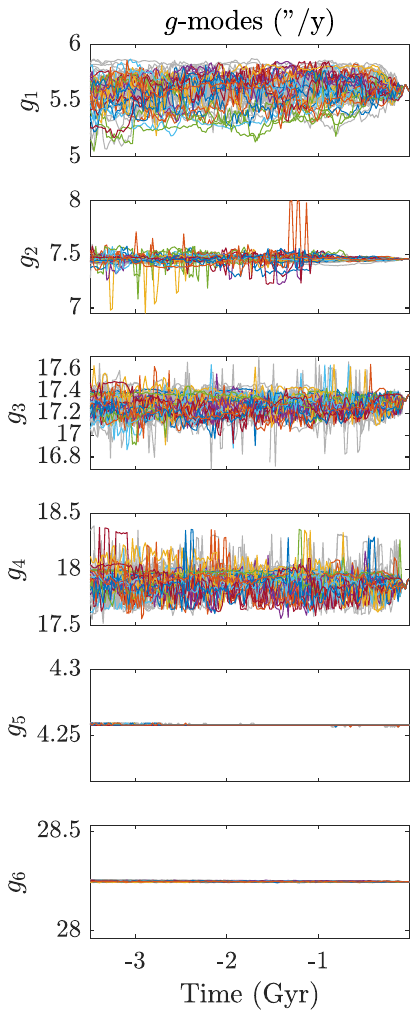}
\vspace*{+00ex} \hspace*{-107ex}
\includegraphics[scale=1.1]{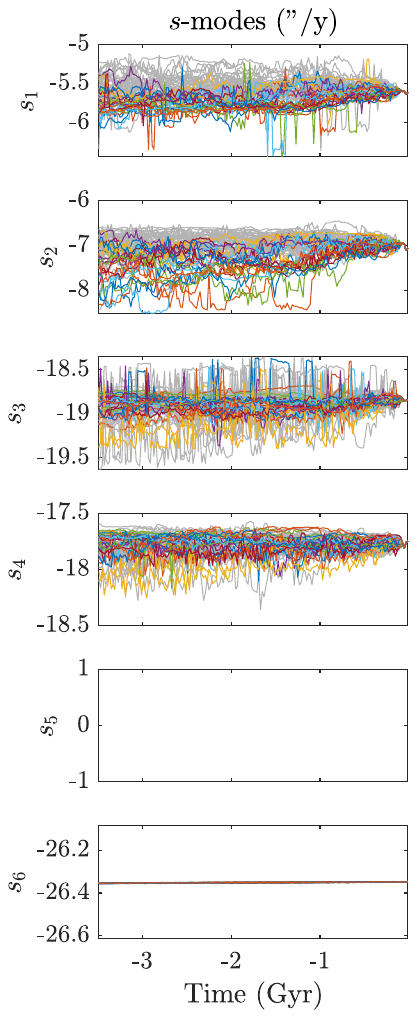}
}
\vspace*{-40ex}
\caption{\scs
Evolution of fundamental (secular) solar system frequencies. 
The $g$- and $s$-frequencies (in arcsec~y\pmo\ = ''~y\pmo)
were determined from our solar system integrations using 
fast Fourier transform (FFT) over consecutive 20-Myr 
intervals and Earth's $k$ and $q$ variables (see text). 
The $g$- and $s$-modes are loosely related 
to the apsidal and nodal precession of the planetary orbits.
Solutions including \sigot-resonance intervals (\otp)
are highlighted in color \citep[see~\ref{sec:lec} and]
[]{zeebelantink24aj}, the remaining solutions are displayed in gray.
\label{fig:gs}
}
\end{figure*}

A 3rd order polynomial fit to observations across the interval 
$t = [-3.5 \ 0]$~Gyr is:
\beqn
(a_L/a_{L0})^{\mbox{\scs fit}} = 1 + q_1 t+ q_2 t^2 + q_3 t^3 \ ,
\label{eqn:poly}
\eeqn
where $\v{q} = [0.1312, \ 0.05197, \ 0.01031]$ and $t$
is in Gyr ($\leq 0$). Importantly, the prescribed fit is 
only used for the OS, not the PT integration, where $a_L$
is calculated using \dtt\ (Fig.~\ref{fig:rr0}a).

\subsection{Luni-solar precession rate}

The luni-solar precession rate $\Psi$ \citep[see e.g.,][]{williams94}
may be calculated from \citep{quinn91}:
\beqn
\Psi = - \dot{\phi} = - d \prc / dt = 
  K(\kap + \bet) \cos \obl + \ggp \ ,
\label{eqn:psi}
\eeqn
where $\prc$ and $\obl$ are the precession and obliquity
angles ($d\prc / dt < 0$, retrograde precession along 
the ecliptic),
$K \ (\kap + \bet) = \alp$ is the precession constant 
(see~\ref{sec:kbet}), and \ggp\ is the geodetic precession (see 
Table~\ref{tab:notval} and \citet{zeebe22aj}).
Note that Eq.~(\ref{eqn:psi}) strictly only applies at $t_0$
and does not capture certain periodic variations.

%
%
%
\section{Results}

\subsection{$g$- and $s$-frequencies}

From our 3.5-Gyr orbital integrations, we determined the solar 
system's fundamental $g$- and $s$-frequencies ($f$'s, aka modes) 
using 
fast Fourier transform (FFT) over consecutive 20-Myr intervals
(Fig.~\ref{fig:gs}). The analysis is straightforward for
practically stable frequencies such as $g_5$, $g_6$, and $s_6$.
However, for full numerical solar system integrations and
nearby, changing frequencies such as $g_3$ and 
$g_4$, and $s_3$ and $s_4$, the analysis is not fail-safe and
cumbersome, and requires individual inspection and manual work.
For example, one may set up an automated search within
a given window for each $f$. However, as $f$'s evolve over time, 
some nearby $f$'s cross into adjacent search windows
($f$ ranges overlap). In addition, some spectral peaks split into 
two peaks across certain time intervals and their power 
varies substantially. Hence the results of an automated search
for $g_3$, $g_4$, $s_3$, and $s_4$ may yield ambiguous
or incorrect results that need to be manually corrected
by adjusting the center and/or width of the search window.
We have checked many but not all of our results
for $g_3$, $g_4$, $s_3$, and $s_4$ displayed in Fig.~\ref{fig:gs}
(total of 64 $\*$ (3,500/20) $\*$ 4 = 44,800 values). Thus,
while the overall patterns for $g_3$, $g_4$, $s_3$, and $s_4$ 
are robust, a few erroneous assignments may be possible.
For $g_1$, $g_2$, $s_1$, and $s_2$ assignment is not 
an issue because their $f$ ranges generally do not overlap.
Values for $s_1~\lsim$$-6.2$~\asy\ (which could interfere
with $s_2$) occurred during so-called \sigot-resonance 
intervals and were confirmed manually. 
The secular resonance $\sigot = (g_1 - g_2) + (s_1 - s_2)$
is dominated by Mercury's and Venus' orbits and can
cause the long eccentricity cycle (LEC) to become 
unstable over long time scales (see~\ref{sec:lec} and 
\citet{zeebelantink24aj}).

The frequencies $g_1$, $s_1$, and $s_2$ drift most strongly 
over time owing to chaotic diffusion. In addition, $g_2$ shows
large and rapid shifts (spikes) at specific times when the 
spectral $g_2$ peak splits into two peaks at significantly reduced
power during \sigot-resonance episodes.
Alternating maximum power
between the two peaks then causes the spikes in $g_2$. As a result, 
$\gtf = (g_2-g_5)$ is unstable and weak/absent during \sigot-resonance
intervals. 
$g_5$, $g_6$, and $s_6$ (dominated by Jupiter and Saturn)
are practically stable over 3.5~Gyr ($s_5$ is zero
due to conservation of total angular momentum/existence
of an invariable plane).

\subsection{Amplitude modulation: $(g_4-g_3)$ and $(s_4-s_3)$}

\begin{figure}[t]
\renewcommand{\baselinestretch}{\fls}
\ifTWO
 \vspace*{-30ex} \hspace*{-16ex}
 \includegraphics[scale=0.6]{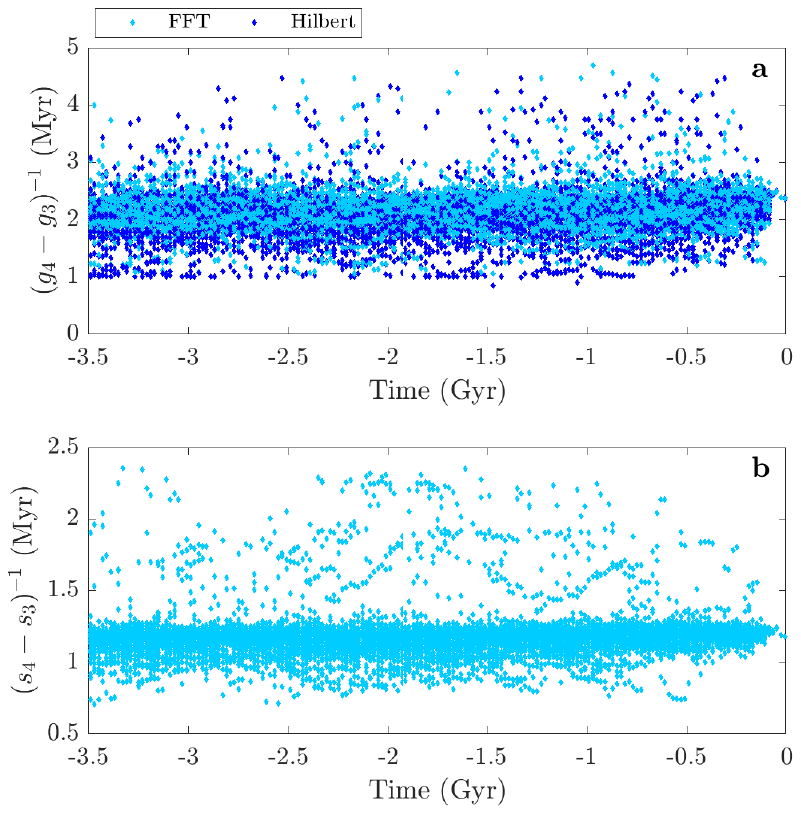}
 \vspace*{-30ex}
\else
 \vspace*{-45ex} \hspace*{-10ex}
 \includegraphics[scale=0.9]{T43.dM3e-4.pdf}
 \vspace*{-40ex}
\fi
\caption{\scs
Periods of $(g_4-g_3)$ and $(s_4-s_3)$ over consecutive 20-Myr 
intervals based on our solar system integrations. Light blue:
from individual $g$'s and $s$'s obtained by direct FFT (see 
Fig.~\ref{fig:gs}). (a) Dark blue: $(g_4-g_3)\pmo~=$~VLEC
from spectral analysis of the
Hilbert transform of a 100-kyr filter of eccentricity
\citep[see Figs. 13 and 14 of][]{zeebe17pa}. For $(s_4-s_3)$,
the Hilbert transform method (using filtered inclination) often
fails because it tends to overemphasize long periods in the
Hilbert transform. For $(s_4-s_3)$ it frequently predicts a 
period doubling (not shown), inconsistent with individual $s$'s 
from FFT.
\label{fig:Tft}
}
\end{figure}

\begin{figure*}[t]
\renewcommand{\baselinestretch}{\fls}
\ifTWO 
 \vspace*{-58ex} \hspace*{+05ex}
 \def\sc{0.70}
 \vspace*{-48ex}
\else
 \vspace*{-60ex} \hspace*{-08ex}
 \def\sc{0.85}
 \vspace*{-60ex}
\fi
\includegraphics[scale=\sc]{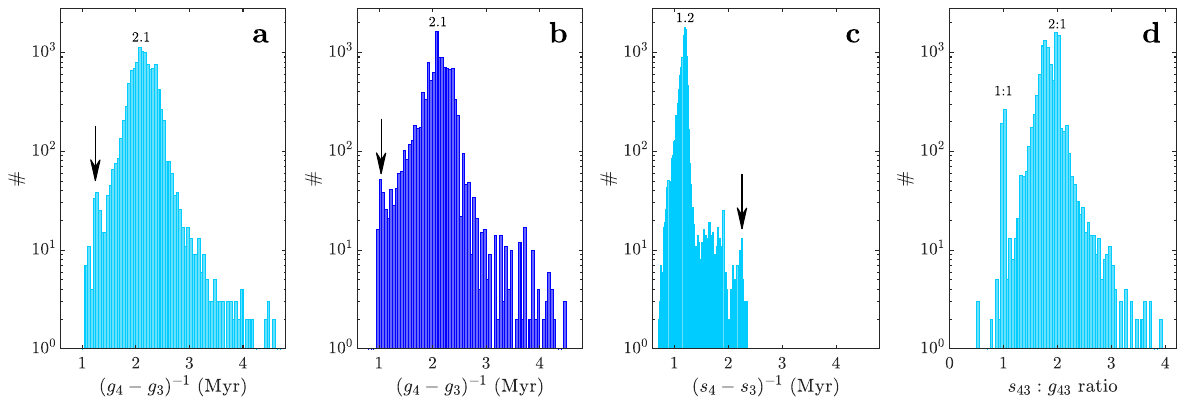}
\caption{\scs
Histograms for the $\gft = (g_4-g_3)$ and $\sft = 
(s_4-s_3)$ periods (see Fig.~\ref{fig:Tft}). Note
the logarithmic $y$-scale.
(a) \gfti\ = VLEC
from direct FFT (max at 2.1~Myr). The side peak 
at \sm{1.25}~Myr (arrow) has been confirmed in various solutions
and contributes to the 1:1 \sft:\gft\ ratio marked in (d).
(b) \gfti\ based on Hilbert transform (max at 2.1~Myr). 
Arrow: same as in (a).
The peak around 3.8~Myr is not robust because the method 
tends to overemphasize long periods in the Hilbert transform.
(c) \sfti\ from direct FFT (max at 1.2~Myr).
The side peak at \sm{2.25}~Myr (arrow) has been found
in a few solutions ($\gfti \simeq~$2.2~Myr simultaneously)
and contributes to the 1:1 \sft:\gft\ ratio as well.
(d) \sft:\gft\ ratio from direct FFT.
\label{fig:hist}
}
\end{figure*}

The frequency combinations $(g_4-g_3) = \gft$ and $(s_4-s_3)
= \sft$ cause amplitude modulations (AM) in eccentricity and 
inclination, respectively, at a period of about 2.37 and 1.18~Myr 
in the recent past. Note that \gft's period is also referred 
to as VLEC.
The frequency ratios are associated with a \tcr{secular} 
resonance (hereafter \sigft) at a recent \gft:\sft\ 
ratio of 1:2, or period ratio of 2:1.
We tested two different methods (FFT and Hilbert 
transform), to determine the periods of \gft\ and 
\sft\ in our 3.5-Gyr integrations, which yielded 
somewhat different results for the period 
distributions
(see Figs.~\ref{fig:Tft} and~\ref{fig:hist}), which, again is
an expression of the fact that the analysis of $g_3$, $g_4$, 
$s_3$, and $s_4$ is not straightforward in all cases (see
above). The most frequent \gft\ period in our 3.5-Gyr 
integrations is \sm{2.1}~Myr (Fig.~\ref{fig:hist}a), 
and not \sm{2.37}~Myr as in the 
recent past. In contrast, the most frequent \sft\ period
is in fact \sm{1.2}~Myr (Fig.~\ref{fig:hist}c),
as in the recent past. As a result,
there is a significant peak in the \sft:\gft\ ratio
distribution at \sm{1.8:1} (2.1/1.2~$\simeq$~1.8, see
Fig.~\ref{fig:hist}d). Nevertheless, in total there are 
more \sft:\gft\ combinations closer to the resonance ratio 
of 2:1 than at 1.8:1.

Importantly, there is a wide 
distribution around the 2:1 ratio (see Fig.~\ref{fig:hist}d),
i.e., the system is not restricted to an exact 2:1 
resonance state. In addition, the \sigft\ \tcr{secular} resonance may 
switch 
to a 1:1 ratio, although we found the 1:1 ratio to be \sm{6}~times 
less frequent than the 2:1 ratio for the 20-Myr intervals
analyzed (Fig.~\ref{fig:hist}d).
Non-integer ratios of \sft:\gft\ (say different from 2:1 and 
1:1) do not only occur while the system transitions between 
the resonance (integer) ratios 2:1 and 1:1. For example, 
in some of our solutions, the system hovers around the 1.8:1 
state for tens of millions of years and returns to the 2:1 
state without transitioning to the 1:1 state in between.
Also, ratios $>$2:1 are common (Fig.~\ref{fig:hist}d)
without transitioning to a potentially higher resonance.
VLEC values (\gft\ periods) anywhere between, 
say 1.2 and 2.8~Myr inferred from deep-time records \citep[e.g.,][]
{olsen19} and \sft:\gft\ ratios between, say 1:1 and 2.5:1 
should not come as a surprise (Fig.~\ref{fig:hist}).

\subsection{Precession-Obliquity evolution}

\begin{figure}[h]
\renewcommand{\baselinestretch}{\fls}
\ifTWO
 \vspace*{-29ex} \hspace*{-16ex}
 \includegraphics[scale=0.59]{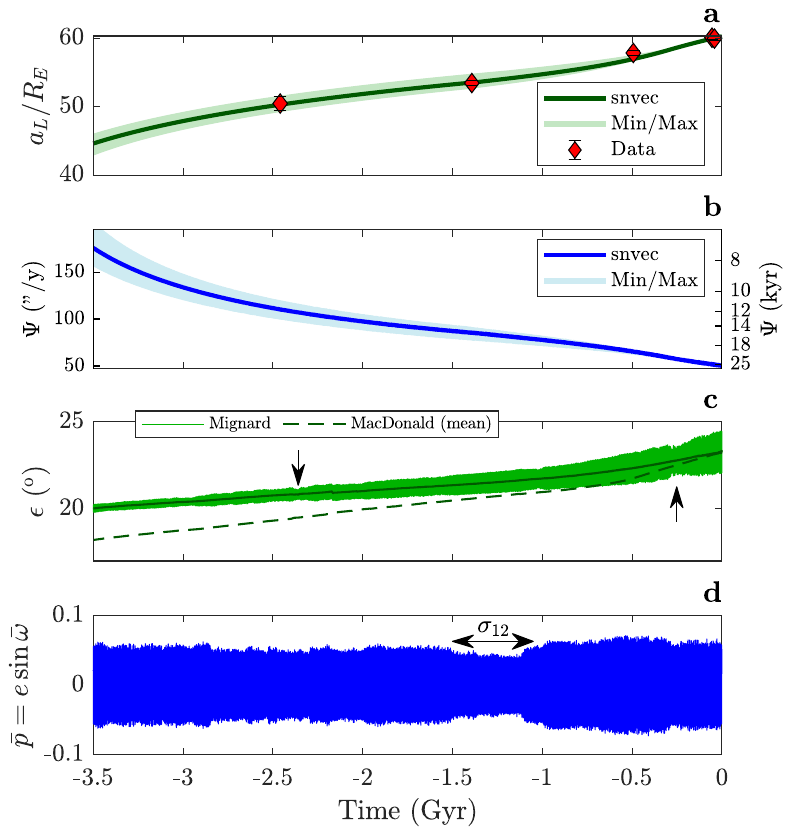}
 \vspace*{-30ex}
\else
 \vspace*{-45ex} \hspace*{-10ex}
 \includegraphics[scale=0.90]{PT.pdf}
 \vspace*{-40ex}
\fi
\caption{\scs
PT solution based on the OS from run R$06$. Error envelopes 
in (a) and (b) are based on estimated lunar distance uncertainties 
(reflecting the cyclostratigraphic data errors, see panel 
(a) and Fig.~\ref{fig:walt}).
(a) Lunar distance ($a_L$) in units of Earth radii ($R_E$).
Green line: PT solution obtained using \snv\ code.
Red diamonds: Observational estimates based on robust data 
sets from cyclostratigraphic studies \tcr{(see Section~\ref{sec:tfpast})}.
(b) Luni-solar precession rate ($\Psi$, see Eq.~(\ref{eqn:psi}))
in arcsec~y\pmo\ (left axis) and period in kyr (right axis).
(c) Obliquity \obl\ calculated with Mignard's torques including
cross-terms (light green) and MacDonald's torques (dashed, 
binned mean values). Arrows highlight selected intervals 
of reduced amplitude variations in orbital inclination and hence 
in \obl. Also note the general trend of increasing \obl\
amplitude with time (see text). 
(d) Climatic precession ($\pbar = e \sin \ombar$).
Reduced amplitude variations in orbital eccentricity (and hence 
in \pbar) may occur during \sigot-resonance episodes (double
arrow, see~\ref{sec:lec} and \citet{zeebelantink24aj}).
\label{fig:PT}
}
\end{figure}

Given a history of the tidal time lag in the past (see 
Fig.~\ref{fig:rr0}) and the OS output 
as input to the PT routine, precession and obliquity can 
be computed over the past 3.5~Gyr for a given OS 
(Fig.~\ref{fig:PT}). The computed evolution of the lunar 
distance ($a_L$, Eq.~(\ref{eqn:dotaL})) and luni-solar 
precession rate ($\Psi$, Eq.~(\ref{eqn:psi})) are nearly the same
for all OS ensemble members. The lunar distance increases
by \sm{35}\% and the luni-solar precession period more than 
triples over 3.5~Gyr (Fig.~\ref{fig:PT}a and~b).
Obliquity (\eps) and climatic precession ($\pbar = e \sin 
\ombar$), however, show differences depending on the OS. 
For example, amplitude variations in Earth's orbital 
inclination (due to OS dynamics) are reflected in \eps\ ---
most evidently intervals of reduced amplitude variation 
(see arrows, Fig.~\ref{fig:PT}c) \citep[cf.][]{zeebe22aj}. 
Similarly, amplitude variations in eccentricity (due to OS 
dynamics) are reflected in \pbar. For instance, the eccentricity
amplitude may be reduced during \sigot-resonance episodes,
which is directly passed on to \pbar\ (see Fig.~\ref{fig:PT}d), 
as eccentricity is the envelope of climatic precession.

The details of the torque physics are relevant for Earth's 
obliquity. Mignard's torques including solar-lunar cross-terms 
\citep{mignard81,touma94} significantly elevate mean \eps\ 
at a given $a_L$ (or time) in the past, relative to MacDonalds's 
torques (Fig.~\ref{fig:PT}c). The cross-terms tend to align 
the spin axis with the orbit normal \citep[see][]{touma94}.
\tcr{Note that an overall lower obliquity in the past
has been discussed in the literature, although the details differ
\citep[e.g.,][]{macdonald64,goldreich66,laskar04NatB,daher21,
farhat22}.}

\begin{figure*}[t]
\renewcommand{\baselinestretch}{\fls}
\ifTWO
 \vspace*{-40ex} \hspace*{+05ex}
 \includegraphics[scale=0.7]{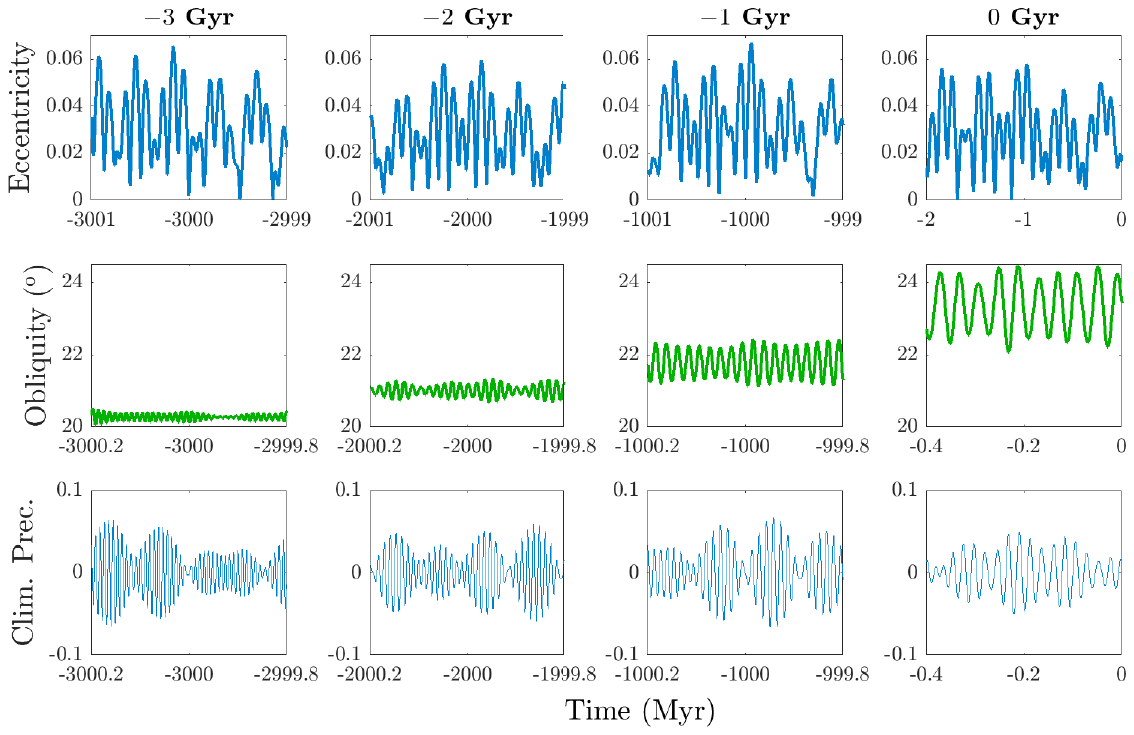}
 \vspace*{-38ex}
\else
 \vspace*{-50ex} \hspace*{-00ex}
 \includegraphics[scale=0.9]{eop.pdf}
 \vspace*{-50ex}
\fi
\caption{\scs
Example of ETP behavior from solution R$02$. 
Top row: Eccentricity over 2-Myr intervals centered around 
$-3, -2, -1$, and $0$~Gyr.
Middle row: Obliquity over 400-kyr intervals.
Bottom row: Climatic precession ($\pbar = e \sin \ombar$) over 
400-kyr intervals. Note that
\pbar's amplitude around $0$~Gyr (bottom-right) is smaller 
than at older time intervals because it happens to coincide with 
a smaller 405-kyr eccentricity maximum in the most recent past 
(see eccentricity, top-right).
\label{fig:eop}
}
\end{figure*}

Importantly, and regardless of the torque details, our 
computations show that Earth's obliquity (\eps) was lower and its 
amplitude (variation around the mean) significantly reduced in 
the past.
The reason for the reduced amplitude is the interplay between 
the luni-solar precession ($\Psi$) and the periodic motion 
of Earth's orbit plane with reference to inclination. Consider
the variations in inclination (secular frequencies $s_i$)
as a forcing acting on the precession motion of the spin
axis (\vs). If $\Psi \gg s_i$, \vs\ closely follows the orbit 
plane (while precessing), \eps\ is constant and the amplitude 
(variation) \sm{null} (past limit). If $\Psi \simeq s_i$, 
the two motions become resonant and \eps's amplitude is large
(future). To first
order, the amplitude (say $B$) is proportional to $s_i/(s_i +\Psi)$
\citep{ward74,ward82}. Thus, the amplitude ratio at two 
different $\Psi$'s (e.g., at $t = -3.5$ and $0$~Gyr) is:
\beqn
\q{B'}{B} = \q{s_i +    \Psi}{s_i   + \Psi'}
     \simeq \q{-18.85 + 50.38}{-18.85 + 176} = 0.2 \ ,
\eeqn
where we used $s_i = s_3 \simeq -18.85$~\asy\ (largest forcing term 
for Earth), $\Psi \simeq 50.38$~\asy\ (at present), and $\Psi' \simeq 
176$~\asy\ (at $-$3.5~Gyr, see Fig.~\ref{fig:PT}b). Given $\max\{B\} 
\simeq 2.4\deg$ over the past 10~Myr, we estimate $\max\{B'\} \simeq 
0.48\deg$ at $-$3.5~Gyr.
Indeed, \eps's maximum amplitude in our numerical integration  
is $\sm{0.45}\deg$ around $-$3.5~Gyr (see Fig.~\ref{fig:PT}c).
Notably, while analyzing the results of the present computations, 
a paper by \citet{ito93} was brought to our attention that
used a different approach but also found low power in obliquity
terms on Gyr-time scale in the past.
In summary, a reduced obliquity amplitude in the past is expected 
from first principles and is quantitatively consistent with our 
numerical solutions (see Discussion).

We include error estimates (envelopes) for our results
based on estimated lunar distance uncertainties (reflecting the 
cyclostratigraphic data errors, see Fig.~\ref{fig:PT}a). 
The uncertainties in 
$a_L$ (minimum/maximum) are propagated through PT integrations
with \snv\ using tidal time lags \dtt's (see Fig.~\ref{fig:rr0}) 
that produce $a_L$ curves coinciding with the lower/upper 
envelope bounds (Fig.~\ref{fig:PT}a). This procedure provides 
propagated errors for $\Psi$ (Fig.~\ref{fig:PT}b) and \obl, 
where the latter also depend on the OS (for details on \obl\ 
errors and a comparison to \citet{waltham15}'s results, 
see~\ref{sec:tiderr}, Fig.~\ref{fig:walt}).

\begin{figure*}[t]
\renewcommand{\baselinestretch}{\fls}
\ifTWO
 \vspace*{-45ex} \hspace*{+05ex}
 \includegraphics[scale=0.7]{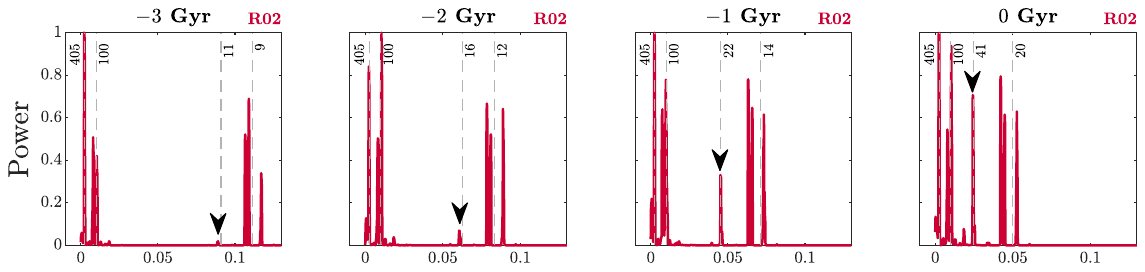}

 \vspace*{-107ex} \hspace*{+05ex}
 \includegraphics[scale=0.7]{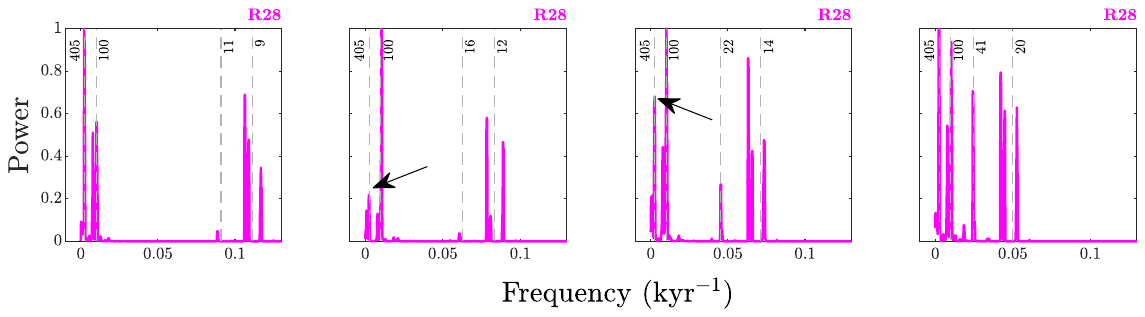}
 \vspace*{-63ex}
\else
 \vspace*{-55ex} \hspace*{-05ex}
 \includegraphics[scale=0.9]{eopMTM-1.pdf}

 \vspace*{-138ex} \hspace*{-05ex}
 \includegraphics[scale=0.9]{eopMTM-2.pdf}
 \vspace*{-87ex}
\fi
\caption{\scs
MTM analysis using 5-Myr windows and an ETP 
composite with relative weights of [1 0.8 0.5] centered 
around $-3, -2, -1$, and $0$~Gyr. Vertical dashed lines 
and numbers indicate periods in kyr.
Top row: solution R$02$. Note rise in obliquity power over 
time (arrows).
Bottom row: solution R$28$. Note substantially reduced LEC 
power at $-2$ and $-1$~Gyr during \sigot-resonance intervals
due to weaker $g_2$ in $g_2-g_5$ (arrows). At the same time, 
the SEC power is reduced around \sm{125-130}~kyr ($g_3-g_2$
and $g_4-g_2$, see Fig.~\ref{fig:AllT}), consistent with a weaker 
$g_2$.
\label{fig:eopMTM}
}
\end{figure*}

\subsection{Example solutions and spectra}

Solar system chaos prevents identifying a unique solution 
on time scales $\gsim$10$^8$~y. Hence we present results
from a few example solutions (and their spectra over 
selected time intervals) that exhibit
typical and/or notable behavior. The solution R$02$
shows a somewhat typical eccen\-tricity-tilt-precession (ETP) 
pattern, without \sigot-resonance intervals (see 
Fig.~\ref{fig:eop}). Note the typical rise in \eps\ and 
its amplitude and the drop in \eps- and \pbar-frequencies
with time. For the spectral analysis (MTM), we selected
5-Myr windows and an ETP composite with relative weights of 
[1 0.8 0.5] centered around $-3, -2, -1$, and $0$~Gyr
(see Fig.~\ref{fig:eopMTM}). The ETP composite was selected so 
that roughly equal power is represented in each frequency 
band (E/T/P) in the most recent past ($0$~Gyr). The MTM
power spectrum of R$02$ illustrates the typical power and
frequency evolution for eccentricity and precession, and 
highlights the common rise in obliquity power over time 
(Fig.~\ref{fig:eopMTM}, top row). 
\tcr{At $t = -1, -2$, and $-3$~Gyr, spectral obliquity power 
based on R$02$ is reduced to about 50\%, 10\%, and 5\% of
its recent value.}
In contrast to R$02$,
the solution R$28$ includes \sigot-resonance intervals 
and shows substantially reduced LEC power at, e.g., $-2$ 
and $-1$~Gyr (Fig.~\ref{fig:eopMTM}, bottom row). For 
details on the unstable LEC and the \sigot\ resonance, 
see~\ref{sec:lec} and \citet{zeebelantink24aj}.

\subsection{Summary of ETP periods}

\begin{figure}[t]
\renewcommand{\baselinestretch}{\fls}
\ifTWO
 \vspace*{-36ex} \hspace*{-25ex}
 \includegraphics[scale=0.7]{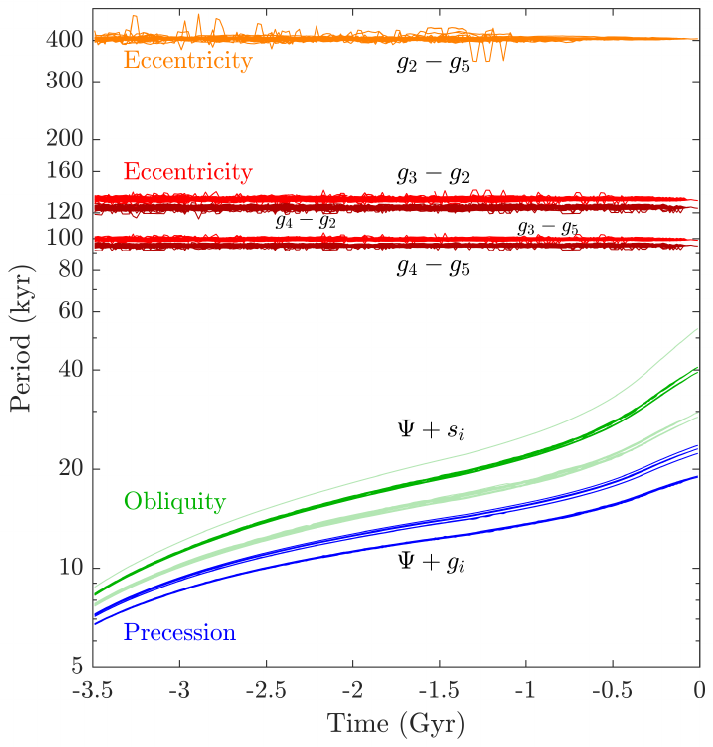}
 \vspace*{-38ex}
\else
 \vspace*{-50ex} \hspace*{-15ex}
 \includegraphics[scale=1.0]{AllTs.pdf}
 \vspace*{-50ex}
\fi
\caption{\scs
Summary of ETP periods from our 3.5-Gyr solutions.
Note the logarithmic ordinate. TP periods are plotted
for the $g_i$'s and $s_i$'s of all OS ensemble members, while 
$\Psi$ was taken from the PT solution based on OS R$06$ 
($\Psi$ is nearly identical for all OSs).
\label{fig:AllT}
}
\end{figure}

From our ensemble integrations, we construct an overview
of the ETP periods over time
(Fig.~\ref{fig:AllT}). The periods of the relevant long- and short 
eccentricity cycles (LEC and SEC) are given by combinations 
of $g_i-g_j$. 
The dominant obliquity frequencies are given by 
$\Psi + s_i$, with $i = 1,\ldots,4,6$. While the $s_i$
vary somewhat over time (see Fig.~\ref{fig:gs}), by far
the largest change in obliquity periods over time
is due to $\Psi$ (see Fig.~\ref{fig:PT}b). A similar
argument holds for the dominant precession frequencies
given by $\Psi + g_i$, with $i = 1,\ldots,5$. The
implications of our findings are discussed in
Section~\ref{sec:disc}.

%
%
%
\section{Discussion \label{sec:disc}}

\subsection{Precession-Obliquity evolution}

Our approach yields a reduced obliquity amplitude in the past 
(Fig.~\ref{fig:PT}c), which is expected from first principles 
and is quantitatively consistent with our numerical solutions
\citep[see also][]{ito93}.
We therefore predict weaker climate forcing at obliquity frequencies 
in deep time and a trend toward reduced obliquity power with age
in stratigraphic records.
In addition, a smaller overall obliquity (polar angle) reduces 
seasonality and would contribute to a muted (high-latitude) 
climate response to obliquity forcing in the distant past.

\subsection{Summary of ETP periods}

\subsubsection{SEC}

Compared to the LEC, the SEC is generally considered a poor 
tuning target in deep time because its exact pattern is unpredictable 
beyond \sm{50}~Myr, as all four SEC components involve either 
$g_3$ or $g_4$ (see Fig.~\ref{fig:AllT}), which are affected 
by solar system chaos on that time scale. In addition, while
the LEC consists of a single component, the SEC consists
of four components, which can rarely, if ever, be extracted 
individually from geologic records. However,
considering that there were likely intervals 
of past LEC ($g_2-g_5$)
instability \citep{zeebelantink24aj}, could the SEC represent
a potential alternative as a primary stratigraphic tuning 
target? 

In our Gyr-simulations, the SEC periods turned out 
to be relatively stable, despite some variations
(Fig.~\ref{fig:AllT}).
Assume for now that the two main SEC pairs (\sm{95-100}~kyr and 
\sm{125-130}~kyr) can be extracted from a stratigraphic record
(usually requires high-quality records).
For a stable tuning target, SEC components
based on combinations including $g_2$ (\sm{125-130}~kyr)
may be less favorable though because of $g_2$'s involvement
in the \sigot\ resonance, which would then leave $g_3-g_5$ and 
$g_4-g_5$ as targets (\sm{95-100}~kyr). As discussed above, 
$g_3$ and $g_4$ do show 
variations in frequency and hence period, whereas $g_5$
is practically stable (see Fig.~\ref{fig:gs}).
More importantly, however, $g_3$ and $g_4$ show 
substantial variations in spectral power. For instance, the 
$g_4/g_3$ power ratio may vary by a factor of 10 to 100 across 
a single 3.5-Gyr solution. As a result, the dominant astronomical
SEC power switches between frequencies over time, which renders
their identification in a record ambiguous. Moreover, if $g_3-g_2$ 
or $g_4-g_2$ is strong during parts of such intervals, and when
time series analysis of the geologic sequence yields only
a single peak around the SEC periods, ambiguities and 
uncertainties in period 
assignment may worsen. For comparison, in our 
simulations, the combined SEC components span a wide range from 
$\min\{g_4-g_5\} \simeq 92$~kyr to
$\max\{g_3-g_2\} \simeq 141$~kyr.
Note that these numbers reflect {\sl individual SEC periods} extracted 
from astronomical solutions, whereas a dominant SEC peak extracted 
from cyclostratigraphic sequences may only reflect an
{\sl average SEC duration} due to the nonlinear response of the 
climate-depositional system and/or insufficient sampling,
spectral resolution, etc.
\citep[see][]{lantink22}. In summary, using the SEC 
as an alternative primary stratigraphic tuning target in deep 
time would likely be challenging in most cases.

\subsubsection{Deep-time obliquity and precession frequencies}

It appears that obliquity and precession frequencies
may be more difficult to distinguish in deep-time stratigraphic
records older than \sm{2.5}~Ga or so (see Fig.~\ref{fig:AllT}). 
However, given the trend of weaker climate forcing at obliquity 
frequencies with age, this might be less of a problem, as 
precession power would dominate over obliquity power (see 
Figs.~\ref{fig:eop} and ~\ref{fig:eopMTM}).
Importantly though, the precession dominance applies to
the orbital forcing, not necessarily the ultimate 
cyclostratigraphic expression, which is generally 
modulated/modified by climate-depositional system.
\tcr{
Also, because our computations predict the most reduced 
obliquity forcing beyond about $-1$~Gyr (Figs.~\ref{fig:eop} 
and ~\ref{fig:eopMTM}), we expect the most significant implications 
for cyclostratigraphic and paleoclimate studies involving 
obliquity for Mesoproterozoic to Archean time intervals.
}

%
%
%
\section{Conclusions}

We have presented internally consistent orbital and 
precession-tilt solutions, including results for
the fundamental (secular) solar system frequencies, 
orbital eccentricity and inclination, lunar distance, 
luni-solar precession rate, Earth's obliquity, and 
climatic precession over the past 3.5~Gyr. Our goal
is to make our theoretical framework widely accessible, 
stimulate computational and observational progress, 
and assist in the interpretation of deep-time 
cyclostratigraphic records and the design of 
external forcing functions in climate models.
Our numerical output is
available at 400-year resolution at \npurl\ and \myurl.
Some of our long-term objectives align closely with Milankovi{\'c},
i.e., ``$\ldots$ be able to determine the most important basic 
features of the Earth's climate computationally.'' \citep{milank41}.
However, we advocate that achieving the objective requires 
a synthesis of both, theoretical and observational efforts.

\vspace*{-3ex}
\section*{Open Research $|$ Data and Software Availability Statement}
{\small
The software associated with this manuscript for the solar system 
integrations (\orb, version 1.0.0) has been published on Zenodo 
\citep{zeebe23aj,zeebe23zn} and GitHub (\giturl).
Data has been deposited at \citet{zeebe24noaa}.
}

\vspace*{0ex}
\noindent
{\small
{\bf Acknowledgments.}
We thank the three reviewers for constructive comments,
which improved the manuscript. We also thank Steve Meyers
for bringing the paper by \citet{ito93} to our attention.
This research was supported by Heising-Simons Foundation 
grants \#2021-2800 and \#2021-2797
(R.E.Z. and M.L.L.) and U.S. NSF grants OCE20-01022, 
OCE20-34660 to R.E.Z. 


\begin{appendix}

{\small
\section{Milankovi{\'c} quote \label{sec:ger}}

In lieu of a footnote (not permitted) we provide here the
original quote cited in Section~\ref{sec:intro} \citep{milank41}:
``Wenn es tats{\"a}chlich gelingen sollte [$\ldots$] eine mathematische 
Theorie zu schaffen, mittels der man die Wirkungen der Sonnenstrahlung
in Raum und Zeit verfolgen k{\"o}nnte, so w{\"a}re man vor allem in der 
Lage, die wichtigsten Grundz{\"u}ge des Erdklimas rechnerisch zu ermitteln.''

\section{Precession constant \label{sec:kbet}}

The calculation of precession and obliquity using our \snv\ code 
is detailed in \citet{zeebe22aj}. A few equations for
the precession constant are included here for completeness.
The change in the spin axis (unit vector \vs) is calculated from 
\citep[e.g.,][]{goldreich66,ward74,ward79,bills90,quinn91,ito95}:
\beqn
\dot{\vs} = \alp (\vn \cdot \vn) (\vs \x \vn) \ ,
\label{eqn:sdot}
\eeqn
where \alp\ is the precession constant and \vn\
the orbit normal (unit vector normal to the orbit plane). 
The obliquity (polar) angle, \obl, is given by:
\beqn
\cos \obl = \vn \cdot \vs \ .
\label{eqn:obl}
\eeqn
The precession (azimuthal) angle, $\prc$, measures the motion
of \vs\ in the orbit plane.
The precession constant \alp\ is calculated from 
\citep{quinn91}:
\beqn
\alp = K \ (\kap + \bet) \ ,
\label{eqn:alp}
\eeqn
where $\kap = (1 - e^2)^{-3/2}$ and $e$ is the orbital eccentricity.
$K$ and \bet\ relate to the torque due to the Sun and Moon,
respectively \citep{quinn91}:
\beqn
K    & = & \q{3}{2} \q{C-A}{C} \q{1}{\om a^3} GM 
\label{eqn:ksun} \\
\bet & = & g_L \ \q{a^3}{a_L^3} \q{m_L}{M} \ ,
\label{eqn:beta}
\eeqn
where $A$ and $C$ are the planet's equatorial and polar moments of 
inertia, $(C-A)/C = E_d$ is the dynamical ellipticity, \om\ is
Earth's angular speed, $a$ the semi-major axis of its orbit,
$a_L$ is the Earth-Moon distance parameter,
and $GM$ is the gravitational parameter of the Sun (see 
Table~\ref{tab:notval}). The index '$L$' refers to lunar properties,
where $g_L$ is a correction factor related to the lunar orbit 
\citep{kinoshita75,kinoshita77,quinn91} and $m_L/M$ is the lunar
to solar mass ratio. The parameter values used for Earth are given
in Table~\ref{tab:notval}. With $K$ and \bet, the luni-solar 
precession rate $\Psi$ can be calculated (see Eq.~(\ref{eqn:psi})).

\section{Numerical integrator and physical setup \label{sec:num}}
Our integrator package {\tt orbitN (v1.0)} \citep{zeebe23aj}
uses a 2nd order
symplectic integrator and Jacobi coordinates \citep{wisdom91,
zeebe15apjA}; the open source code is available at \zenurl\ and 
\giturl. Solar system integrations were carried out following
our earlier work for which methods and integrator have been 
extensively tested and compared against other studies
\citep{zeebe15apjA,zeebe15apjB,zeebe17aj,zeebelourens19,zeebe22aj,
zeebe23aj}.
For the present study, we also included simulations with 
an independent integrator package (HNBody) \citep{rauch02}
and found essentially the same dynamical behavior.
All simulations include contributions from general 
relativity, available in \orb\ as Post-Newtonian effects 
due to the dominant mass \tcr{\citep{saha94}.}
The Earth-Moon system was modeled as a gravitational 
quadrupole \citep{quinn91,varadi03,zeebe17aj,zeebe23aj}.
Initial conditions for the positions and velocities of 
the planets and Pluto were generated from the 
latest JPL ephemeris DE441 \citep{park21de}
using the SPICE toolkit for Matlab. 
Coordinates were obtained at JD2451545.0 
in the ECLIPJ2000 reference frame and subsequently 
rotated to account for the solar quadrupole moment 
($J_2$) alignment with the solar rotation axis 
\citep{zeebe17aj}. Solar mass loss was included
using $\dot M/M = -7 \x 10^{-14}$~y\pmo\ 
\citep[e.g.,][]{quinn91}. 
\tcr{Quinn at al.'s value falls toward the lower end of 
more recent solar mass loss estimates \citep[e.g.,][]{minton07,
fienga24}, which is likely of minor importance though.
In additional eight test simulations (not shown), we used a very 
large solar mass loss of $\dot M/M = -1.1 \x 10^{-11}$~y\pmo\ 
\citep{spalding18} and found again essentially the same dynamics, 
including \sigot-resonance episodes. One consequence of
the large mass loss is of course a substantial secular 
trend in fundamental frequencies \citep[see][]{spalding18}.
}
As solar mass loss also causes a 
secular drift in total energy, we added test runs 
with $M = const.$ to check the integrator's numerical 
accuracy. Total energy and angular momentum errors were 
small throughout the 3.5-Gyr integrations (relative errors: 
$\lsim$$6 \x 10^{-10}$ and $\lsim$$7 \x 10^{-12}$, 
see \citet{zeebelantink24aj}).
\tcr{The current simulations did not consider asteroids,
which is numerically expensive when included as fully
gravitationally interacting bodies.
Asteroids are quite important for, e.g., high-fidelity solutions 
on 100-Myr time scale, when probing for sensitivities to 
dynamical chaos. However, the current goal is to explore the 
general solution/phase space of the system and provide
characteristic features of long-term forcing and
fundamental frequencies, which is unlikely to be affected
by asteroids due to their small mass.
}
Our default numerical timestep ($|\D t| = 4$~days) is close to the 
previously suggested value of 3.59~days to sufficiently 
resolve Mercury's perihelion \citep{wisdom15,hernandez22,abbot23}.
In additional eight simulations, we tested $|\D t| = 2.15625$~days
(adequate to $\eM \ \lsim 0.4$) and found no differences
in the results, including in terms of \sigot\ resonances
(see~\ref{sec:lec} and \citet{zeebelantink24aj}), which occurred 
in 3/8 solutions.

\section{Long eccentricity cycle and \sigot\ resonance \label{sec:lec}}

As detailed in \citet{zeebelantink24aj}, we found
that the long eccentricity cycle (LEC) can become unstable 
over long time scales, without major changes in, or destabilization 
of, planetary orbits. The LEC's disruption is due to a secular 
resonance between the apsidal and nodal precession frequencies 
dominated by Mercury's and Venus' orbits and is a major contributor 
to solar system chaos. Entering/exiting the secular resonance is 
a common phenomenon on long time scales, occurring in \sm{40}\% of 
our astronomical solutions. During resonance episodes, the LEC 
is very weak or absent and Earth's orbital eccentricity and 
climate-forcing spectrum are unrecognizable compared to the recent past. 
These findings have fundamental implications for paleoclimatology, 
astrochronology, and cyclostratigraphy because the paradigm that 
the longest Milankovi{\'c} cycle dominates Earth's climate forcing, 
is stable, and has a period of \sm{405}~kyr requires revision
\citep{zeebelantink24aj}.

\section{Tidal torques \label{sec:trqs}}

\subsection{Mignard's torques}

As our default option, we use Mignard's tidal torques, including 
solar-lunar cross terms \citep{mignard81,touma94}. The equations 
can be written succinctly using abbreviations following 
\citet{goldreich66}:
\ifTWO
\beqn
x = (\vs \* \vn) = \cos \eps  ; &
y = (\vb \* \vn) = \cos i_L   ; & \nonumber \\ &
z = (\vs \* \vb) = \cos \epss               \\
s_x = (1-x^2)^\q{1}{2} = \sin \eps  ; &
s_y = (1-y^2)^\q{1}{2} = \sin i_L   ; & \nonumber \\ &
s_z = (1-z^2)^\q{1}{2} = \sin \epss \ .
\eeqn
\else
\beqn
x = (\vs \* \vn) = \cos \eps  \quad ; \quad &
y = (\vb \* \vn) = \cos i_L & \quad ; \quad
z = (\vs \* \vb) = \cos \epss               \\
s_x = (1-x^2)^\q{1}{2} = \sin \eps  \quad ; \quad &
s_y = (1-y^2)^\q{1}{2} = \sin i_L & \quad ; \quad
s_z = (1-z^2)^\q{1}{2} = \sin \epss \ .
\eeqn
\fi
Furthermore, we use ($a_E =$ semimajor axis of Earth's orbit):
\beqn
A_1    = \dtt \q{\ktt G m_L^2 R_E^5}{a_L^6} \quad ; \quad
A_2    = \dtt \q{\ktt G M^2   R_E^5}{a_E^6} \quad ; \quad
\ifTWO \nonumber \\ \fi
A_{12} = \dtt \q{\ktt G m_L M R_E^5}{a_L^3 a_E^3}
\eeqn
and
\beqn
u   = \cos  (\Om - \Om_{\sun}) = (y - xz) / (s_x s_z) \quad ; \quad
\ifTWO \nonumber \\ \fi
u_2 = \cos 2(\Om - \Om_{\sun}) = 2u^2 - 1 \ ,
\eeqn
where $\Om$ and $\Om_{\sun}$ are the ascending nodes of 
the lunar and solar orbits on the equator plane. The
torque components in Goldreich's coordinate projection 
are given by:
\beqn
\vT \* \vs & = & T_2 s_z   + T_3 z            + T'_2 s_x          + T'_3 x \\
\vT \* \vb & = & T_3       - T'_1 w/s_x       + T'_2 (z - xy)/s_x + T'_3 y \\
\vT \* \vn & = & T_1 w/s_z + T_2 (x - yz)/s_z + T_3 y             + T'_3 \ .
\eeqn
Following \citet{goldreich66} and \citet{touma94}, we
assume negligible contributions from $T_1$ and $T'_1$
terms and drop them. Using $f_2  = (x - yz)/s_z$ and 
$f'_2 = (z - xy)/s_x$, the components $T_i$ and $T'_i$ can
be easily identified (see below). The torques below are given
for the Moon and are opposite for the Earth. Then, the 
average lunar torques due to lunar tides can be written as:
\begin{alignat}{3}
\vT \* \vs & =
    A_1 \left[
        \q{3}{2} \om s_z^2 + 3 z (\om z - n_L)
        \right] && = T_2 s_z + T_3 z \\
\vT \* \vb & =
    A_1 \left[
        3 (\om z - n_L)
        \right] && = T_3 \\
\vT \* \vn & =
    A_1 \left[
        \q{3}{2} \om (x - yz) + 3 y (\om z - n_L)
        \right] && = T_2 f_2 + T_3 y \ .
\end{alignat}
The average solar torques due to solar tides are:
\begin{alignat}{3}
\vT \* \vs & =
    A_2 \left[
        \q{3}{2} \om s_x^2 + 3 x (\om x - n_E)
        \right] && = T'_2 s_x + T'_3 x \\
\vT \* \vb & =
    A_2 \left[
        \q{3}{2} \om (z - xy) + 3 y (\om x - n_E)
        \right] && = T'_2 f'_2 + T'_3 y \\
\vT \* \vn & =
    A_2 \left[
        3 (\om x - n_E)
        \right] && = T'_3 \ .
\end{alignat}
The average lunar torques due to solar tides (cross terms)
are:
\begin{alignat}{3}
\vT \* \vs & =
 A_{12} \om \left[
        \q{3}{8} s_y^2 s_x^2 u_2
       -\q{9}{8} s_y^2 s_x^2
       -\q{3}{4} x y s_x s_y u 
       +\q{3}{4} s_x^2
        \right]  && = T'_2 s_x + T'_3 x \\
\vT \* \vb & = 0 && \\
\vT \* \vn & =
 A_{12} \om \left[
       -\q{3}{4} y s_x s_y u 
        \right]  && = T'_3 \ .
\end{alignat}
Finally, the average solar torques due to lunar tides (cross terms)
are:
\begin{alignat}{3}
\vT \* \vs & =
 A_{12} \om \left[
        \q{3}{8} s_y^2 s_x^2 u_2
       -\q{9}{8} s_y^2 s_x^2
       -\q{3}{4} x y s_x s_y u 
       +\q{3}{4} s_x^2
        \right]  && = T'_2 s_x + T'_3 x \\
\vT \* \vb & =
 A_{12} \om \left[
        \q{3}{4} y^2 s_x s_y u
       -\q{3}{4} x y s_y^2
        \right]  && \\
\vT \* \vn & = 0 \ . &&
\end{alignat}

\subsection{MacDonald's torques}

\def\nsp{\!\!\!\!\!}
MacDonald's torques are given by
\citep{macdonald64,goldreich66} ($T_1 = 0$):
\beqn
T_2  \nsp & = & \nsp -T_{2L} =
           -\left[ 2 \ m_L \ A / (\pi  a_L^6) \right] \ q  B(q^2) \sin(2\del)  
\label{eqn:Tii} \\
T_3  \nsp & = & \nsp -T_{3L}  =
           -\left[ 2 \ m_L \ A / (\pi  a_L^6) \right] \ q' K(q^2) \sin(2\del)
\label{eqn:Tiii} \\
A   \nsp & = &  \nsp (3/2) \ G \ m_L \ R_E^5 \ \ktt \ ,
\eeqn
where $T_i$ and $T_{iL}$ refer to the Earth and the Moon, 
respectively, and \del\ is the tidal phase lag.
Furthermore,
\beqn
q^2 = \q{1 - z^2}{1 + \tldalp^2 -2 \tldalp z} \ ,
\eeqn
where
$z = (\vs \* \vb) = \cos \epss$, $\tldalp = n_L / \om$, and
$q' = (1 - q^2)^{1/2}$, with sign $q'$ = sign 
$(z - \tldalp)$. The complete elliptic  integrals of the 1st 
and 2nd kind are denoted as $K(q^2)$ and $E(q^2)$, from which 
$B(q^2)$ is calculated:
\beqn
B(q^2) = \{ E(q^2) - q'^2 K(q^2) \} / q^2 \ .
\eeqn

\section{Tidal evolution and error envelopes \label{sec:tiderr}}

In this appendix, we provide additional information on
tidal evolution models (Fig.~\ref{fig:PTF22}) and a 
comparison of our computed orbital parameters to 
\citet{waltham15}, including error envelopes 
(Fig.~\ref{fig:walt}).

\begin{figure}[h]
\renewcommand{\baselinestretch}{\fls}
\ifTWO
 \vspace*{-33ex} \hspace*{-18ex}
 \includegraphics[scale=0.65]{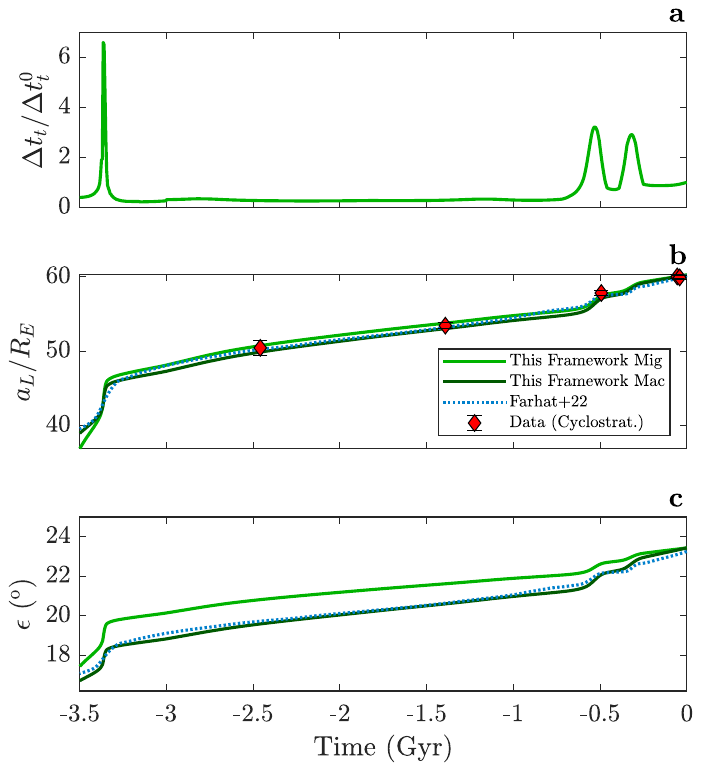}
 \vspace*{-25ex}
\else
 \vspace*{-60ex} \hspace*{-15ex}
 \includegraphics[scale=1.0]{PTF22.pdf}
 \vspace*{-40ex}
\fi
\caption{\scs
Illustration of evaluating results of ocean tidal evolution models
within the current framework. As an example, we select the recent
model by \citet{farhat22}. The point here is to illustrate
the procedure, not to reproduce exact results. 
(a) Tidal time lag (normalized to \dtt\ at $t = 0$) used here in 
Mignard's and MacDonald's torques (Mig and Mac), 
roughly approximated
to resemble \citet{farhat22}'s tidal torque history
and to match $a_L$ data.
(b) Lunar distance ($a_L$) in units of Earth radii ($R_E$).
Green line: computed using current approach and
time lag shown in (a). Dashed blue line: results from 
\citet{farhat22}.
Red diamonds: Observational estimates based on robust data 
sets from cyclostratigraphic studies 
\tcr{(see Section~\ref{sec:tfpast})}.
(c) Earth's obliquity \obl. Larger \obl\ values at a given 
$a_L$ (Mignard's torques) are due to solar-lunar cross-terms 
(see main text). Otherwise, and for the most part, there are 
small differences to the current standard approach
(see Fig.~5). Importantly, no observational 
data is presently available 
to verify or falsify the rapid rise in lunar distance around 
$-$3.5~Gyr, see (b). For Earth's obliquity (c), the tidal torque 
physics are more important here than the details of the tidal history, 
as long as $a_L$ matches observational data (fitted here using 
\dtt).
\label{fig:PTF22}
}
\end{figure}

\begin{figure}[p]
\renewcommand{\baselinestretch}{\fls}
\ifTWO
 \vspace*{-28ex} \hspace*{-22ex}
 \includegraphics[scale=0.65]{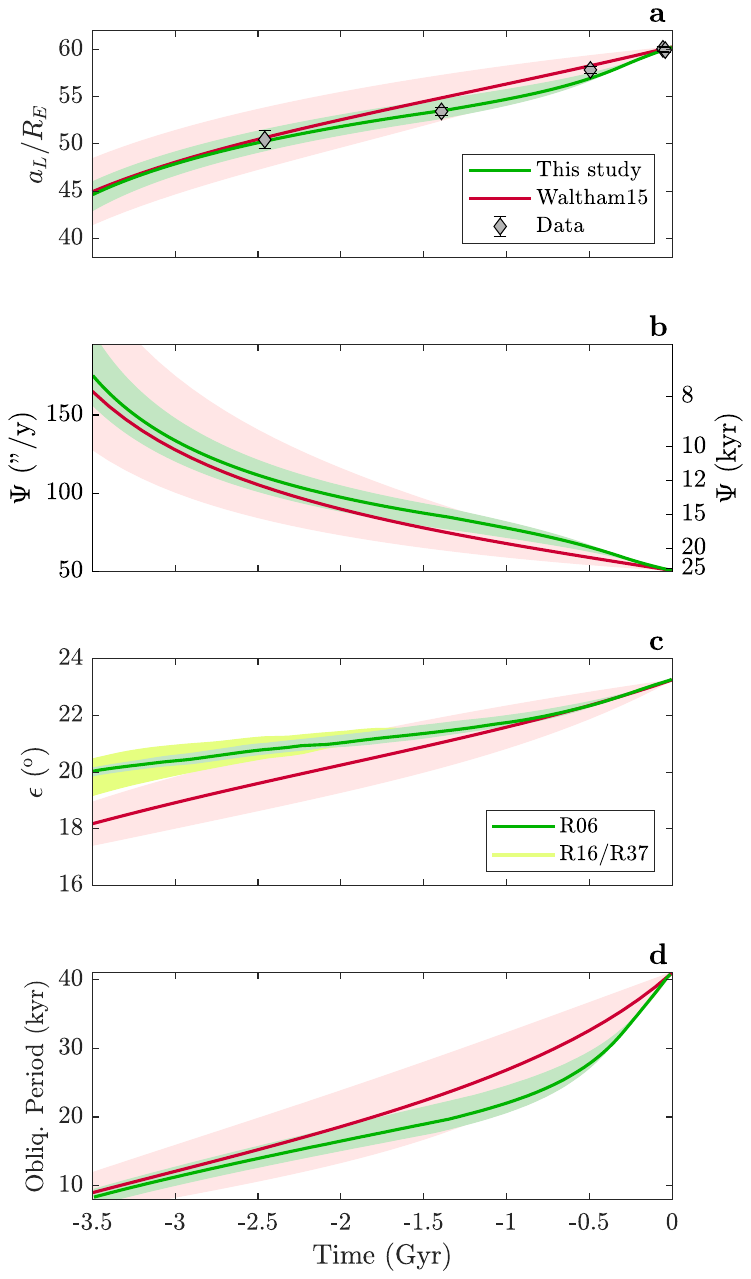}
 \vspace*{-18ex}
\else
 \vspace*{-18ex} \hspace*{-10ex}
 \includegraphics[scale=0.9]{Walt15.pdf}
 \vspace*{-22ex}
\fi
\caption{\scs
Comparison of the present results (based on OS R$06$) to those 
of \citet{waltham15},
see \walturl. \citet{waltham15} used a simplified model which 
neglects solar and solar-lunar cross tides.
Light red areas: Error envelope given by \citet{waltham15}. 
Light green areas: Error envelope of this study based on 
estimated lunar distance uncertainties (reflecting the 
cyclostratigraphic data errors, see panel (a)).
(a) Lunar distance ($a_L$) in units of Earth radii ($R_E$).
Gray diamonds: Observational estimates based on robust data 
sets from cyclostratigraphic studies
\tcr{(see Section~\ref{sec:tfpast})}.
(b) Luni-solar precession rate ($\Psi$)
in arcsec~y\pmo\ (left axis) and period in kyr (right axis).
(c) Earth's obliquity \obl. To facilitate the comparison, 
the \obl\ results of this study (which include the full
\obl\ amplitude, i.e., variations around the mean) were binned and 
smoothed.
Green-yellow area: Extended error envelope including OSs R$16$ and 
R$37$ (smallest and largest binned \obl\ values of ensemble 
at $t = -3.5$~Gyr).
Note that different OSs have little effect on $a_L$ and $\Psi$.
Larger \obl\ values at a given time in the past \citep[compared 
to, and significantly different from,]
[]{waltham15} are calculated here using Mignard's torques and
are hence due to solar-lunar cross-terms \citep[see][]{touma94}.
(d) Obliquity period.
\label{fig:walt}
}
\end{figure}

} 

\end{appendix}


\ifTWO
 \renewcommand{\baselinestretch}{0.0}
\else
 \clearpage
 \newpage
 \renewcommand{\baselinestretch}{0.8}
\fi





\end{document}